\documentclass[%
 reprint,
 bibnotes,
 amsmath,amssymb,
 aps,
]{revtex4-1}

\usepackage{graphicx}
\usepackage{dcolumn}
\usepackage{bm}
\usepackage{multibib}
\usepackage[section]{placeins}
\newcommand\Tstrut{\rule{0pt}{2.6ex}}         

\begin{document}

\preprint{APS/123-QED}

\title{Intermolecular correlations are necessary to explain diffuse scattering from protein crystals}

\author{Ariana Peck$^1$}
\email[Correspondence to: ]{apeck@stanford.edu.}
\author{Fr{\'e}d{\'e}ric Poitevin$^{2,3}$} 
\author{Thomas J. Lane$^{4}$}
\affiliation{$^1$Department of Biochemistry, Stanford University, Stanford, CA, 94305, United States.\\
$^2$Department of Structural Biology, Stanford University, Stanford, CA 94305, United States.\\
$^3$Stanford PULSE Institute, SLAC National Accelerator Laboratory, Menlo Park, CA 94025, United States\\
$^4$Bioscience Division and Linac Coherent Light Source, SLAC National Accelerator Laboratory, Menlo Park, CA 94025, United States}

\date{\today}

\begin{abstract}
Conformational changes drive protein function, including catalysis, allostery, and signaling. X-ray diffuse scattering from protein crystals has frequently been cited as a probe of these correlated motions, with significant potential to advance our understanding of biological dynamics. However, recent work challenged this prevailing view, suggesting instead that diffuse scattering primarily originates from rigid body motions and could therefore be applied to improve structure determination. To investigate the nature of the disorder giving rise to diffuse scattering, and thus the potential applications of this signal, a diverse repertoire of disorder models was assessed for its ability to reproduce the diffuse signal reconstructed from three protein crystals. This comparison revealed that multiple models of intramolecular conformational dynamics, including ensemble models inferred from the Bragg data, could not explain the signal. Models of rigid body or short-range liquid-like motions, in which dynamics are confined to the biological unit, showed modest agreement with the diffuse maps, but were unable to reproduce experimental features indicative of long-range correlations. Extending a model of liquid-like motions to include disorder across neighboring proteins in the crystal significantly improved agreement with all three systems and highlighted the contribution of intermolecular correlations to the observed signal. These findings anticipate a need to account for intermolecular disorder in order to advance the interpretation of diffuse scattering to either extract biological motions or aid structural inference.
\end{abstract}

\maketitle


\section{\label{sec:introduction}Introduction}

X-ray diffraction images from macromolecular crystals frequently exhibit a diffuse background between and beneath Bragg peaks \cite{pmid24507780, welberry2016}. In contrast to the Bragg reflections which arise from coherent diffraction across the crystal, this diffuse signal results from disorder-induced incoherent diffraction. The uncorrelated disorder of solvent and macromolecular atoms yields a trivial diffuse scattering pattern that is radially symmetric \cite{pmid19836331}. On the other hand, correlated disorder produces anisotropic diffuse scattering features whose spacing and intensity in reciprocal space are respectively determined by the length scale and amplitudes of the correlated atomic displacements involved \cite{pmid7568674}.

Correlated displacements in macromolecules underlie many biological functions, such as allostery, signaling, and enzyme catalysis. However, methods for directly measuring such motions with high spatial resolution are rare. Diffuse scattering has routinely been cited as one such method that could provide unique insights into the collective motions responsible for biological functions \cite{pmid24507780, pmid7568674}. Indeed, this has motivated the majority of studies of diffuse scattering from macromolecular crystals \cite{pmid9438860, pmid9636718, pmid15299461, pmid15681654, pmid1600083, pmid16384188, pmid17154425, pmid25453071, pmid27035972}, despite the technical challenges of measuring and the computational cost of modeling this signal \cite{pmid24507780}.

Recently, the application of diffuse scattering to complement and improve (static) structure inference from crystal diffraction has also been proposed. Chapman and colleagues have suggested that in cases where measurable diffuse scattering extends to higher resolution than the Bragg data, it may be employed for structural inference at that higher resolution \cite{pmid26863980}. Further, the ability to oversample the diffraction pattern by measuring the continuous diffuse signal raises the possibility of solving the phase problem directly, without resorting to anomalous or isomorphic methods. This approach, however, assumes that diffuse scattering primarily originates from specific types of disorder, such as rigid body motions, which are unlikely to inform on biological function.

Identifying the physical origins of diffuse scattering, and thus its potential for probing biological motions or advancing methods, remains a challenge for the field. Many types of disorder involve small motions that can be conveniently described by a matrix, whose elements give the covariation between any two atoms' displacements from their mean positions. Such a covariance matrix can directly, but not uniquely, predict diffuse scattering. If the covariance matrix could be inferred directly from experiment, the diffuse signal could be analyzed to determine which regions of the macromolecule move together. There is a fundamental problem with direct inference, however: while the number of observed independent variables is quite large (say, $V$ voxels for a given unit cell volume, assuming the maximum resolution of diffraction is fixed), the number of unknowns is even larger (order $V^2$ matrix elements, one for each atom pair, assuming the number of atoms scales linearly with volume). Thus, to infer a covariance matrix one must make simplifying assumptions about the nature of protein motions: a parsimonious model for the protein physics is required. 

In this work, we analyze the parsimonious models that have been suggested previously (reviewed in \cite{pmid28558231}) and variations of these models to critically examine the types of disorder underlying the diffuse scattering observed in a range of systems. The three protein systems analyzed here represent both diverse crystalline properties and biological functions: cyclophilin A (CypA), a monomeric proline isomerase \cite{pmid27035972}; WrpA, a flavodoxin-like protein \cite{pmid27189921}; and a dimeric enzyme, alkaline phosphatase (AP), bound to its transition state analog \cite{pmid26858101}. Despite this diversity, we found that multiple models of intramolecular conformational dynamics were unable to explain the observed diffuse scattering in all three systems. By contrast, models of simpler dynamics, including rigid body and liquid-like motions, consistently showed modest correlation with the experimental signal when correlations were confined to the biological unit. The conventional form of the liquid-like motions model, in which correlations span neighboring protein molecules in the crystal, significantly improved the agreement both quantitatively and qualitatively, but still did not fully account for the observed signal. This comprehensive comparison both reconciles opposing viewpoints about the principal physical origins of diffuse scattering and specifically identifies intermolecular correlations as a critical component of the underlying disorder. Further, these findings anticipate that deconvolving the intermolecular contribution from the signal will be required to enable the future application of diffuse scattering to probe biological motions or improve static structure inference.

\section{Methods}

\begin{table*}
\caption{\label{tab:stats}Data collection and statistics}
\begin{ruledtabular}
\begin{tabular}{lccc}
& CypA & AP & WrpA \\ \hline
\textit{Data Collection and Bragg statistics \footnote{Values in parenthesis are for the highest resolution shell.}} & \Tstrut \\ \hline 
Space group & P2$_1$2$_1$2$_1$ & P6$_3$22 & P4$_2$22  \Tstrut \\
Unit cell axes &  \\
\hspace{4pt} \textit{a, b, c} (\AA) & 42.9 52.4 89.1 & 161.3, 161.3, 139.4 & 61.3 61.3 128.7 \\
\hspace{4pt} \textit{$\alpha$, $\beta$, $\gamma$} ($^{\circ}$) & 90.0 90.0 90.0 & 90.0, 90.0, 120.0 & 90.0 90.0 90.0 \\
Wavelength (\AA) & 0.9795 & 0.9795 & 0.9787 \\
Oscillation range ($^{\circ}$) & 0.5 & 0.15 & 1.0 \\
Collection temperature (K) & 273 & 100 & 100 \\
Beam divergence S.D. ($^{\circ}$) & 0.0417 & 0.0285 & 0.0339 \\
Mosaicity ($^{\circ}$) & 0.106 & 0.136 & 0.197 \\
Resolution range (\AA) & 44.57 - 1.20 (1.24 - 1.20) & 49.36 - 2.00 (2.07 - 2.00) & 44.38 - 2.51 (2.60 - 2.51) \\
Multiplicity & 5.8 (4.8) & 6.3 (6.5) & 12.7 (12.7) \\
Completeness (\%) & 93.6 (85.8) & 94.0 (92.7) & 99.6 (98.5) \\
$\langle$I/$\sigma$I$\rangle$ & 11.3 (3.9) & 26.2 (3.6) & 25.5 (4.6)  \\
R$_\text{merge}$ & 0.113 (0.504) & 0.045 (0.348) & 0.061 (0.520) \\ 
R$_\text{meas}$ & 0.123 (0.564) & 0.049 (0.375) & 0.063 (0.542) \\ 
CC$_{1/2}$ & 0.989 (0.872) & 1.000 (0.972) & 0.999 (0.962)  \\ 
Wilson B (\AA$^2$) & 16.1 & 31.4 & 64.7 \\
Detector & PILATUS 6M & PILATUS 6M & MAR Mosaic 300 \\ 
SBGrid ID & 68 & 456 & 203 \\ \hline 
\vspace{2pt} \textit{Structure refinement statistics} & \Tstrut \\ \hline
Resolution range (\AA) & 44.57 - 1.20 & 49.36 - 2.00 & 44.37 - 2.50 \Tstrut \\
No. of non-hydrogen atoms & 1419 & 6731 & 1237 \\
Average B-factor (\AA$^2$) & 21.4 & 46.2 & 77.0 \\
Solvent content (\%) & 56 & 62 & 52 \\
R$_\text{work}$/R$_\text{free}$ & 0.169/0.177 & 0.192/0.231 & 0.235/0.277 \\ \hline
\vspace{2pt} \textit{Diffuse scattering statistics} & \Tstrut \\ \hline
Resolution range (\AA) & 93.4 - 1.5 & 75.2 - 2.5 & 69.3 - 2.1 \Tstrut \\
CC$_\text{Friedel pairs}$\footnote{For values in parenthesis, the average radial intensity was subtracted prior to applying symmetry operations.} & 0.979 (0.623) & 0.954 (0.695) & 0.988 (0.689) \\
CC$_\text{unsym., Friedel-sym.}$$^\text{b,}$$\footnote{Correlation coefficient between the unsymmetrized map and the map after applying the indicated symmetry operations.}$ & 0.996 (0.915) & 0.994 (0.957) & 0.997 (0.930) \\
CC$_\text{unsym., Laue-sym.}$$^\text{b,c}$ & 0.993 (0.862) & 0.984 (0.887) & 0.994 (0.827) \\
Completeness (\%) & 98.7 & 96.1 & 100.0 \\
\end{tabular}
\end{ruledtabular}
\end{table*}

\subsection{Reconstruction of 3d diffuse scattering maps}
For each experimental dataset, Bragg reflections were indexed by XDS \cite{pmid20124692}. Refined parameters from XDS were then used to determine the reciprocal space coordinates of each pixel in every diffraction image. Measured intensities were corrected for polarization of the x-ray beam \cite{doi:10.1063/1.1319614} and the difference in solid angle subtended by pixels at different scattering angles \cite{pmid19488705}. Per-image scale factors from XDS were applied to correct for differences in overall intensity across the rotation range \cite{pmid20124692}. For datasets collected on a Pilatus detector, parallax broadening was also accounted for, as implemented in DIALS \cite{pmid27050135}. No Lorentz correction was applied, as pixel intensities were averaged rather than integrated, and the diffuse features were observed to vary slowly in reciprocal space. Under these conditions, the volume of reciprocal space integrated by each pixel needs only to be corrected for the solid angle subtended, and not the arc length the pixel traverses due to rotation of the crystal \cite{Boysen}.

After correcting for geometrical distortions, Bragg peaks were removed by implementing the spot prediction algorithm described in Ref. \cite{pmid20124693}. Pixels predicted to be spanned by a Bragg reflection were masked if their intensity exceeded three standard deviations above the mean intensity of neighboring pixels outside the predicted reflection region in a 30 x 30 pixel window centered on the reflection. Prior studies have replaced masked pixel intensities by the intensities from adjoining pixels \cite{pmid26457426} or pre-filtered images \cite{pmid26863980}, but the strategy of masking without replacement was found to maintain an adequate signal-to-noise ratio for map voxels that coincided with Miller indices (Fig. S1, cf. solid and dashed lines). To ensure the complete removal of Bragg contaminants, an additional step of masking was performed to eliminate pixels whose intensities exceeded the median radial intensity by more than five times the median absolute deviation of that resolution shell. In the case of WrpA, pixels were additionally masked if they fell within a region contaminated by scattering from the cryo-loop and exceeded the median radial intensity by more than 2.5 times the median absolute deviation of the resolution shell. Fig. S2 shows the results of this Bragg masking procedure for a representative image from CypA.

The images' radial intensity profiles were then compared for uniformity to ensure that independent measurements of equivalent positions in reciprocal space were on the same scale prior to merging. As noted above, variation in the total protein scattering across the rotation range, due to fluctuations in the illuminated crystal volume and incident beam intensity for instance, was corrected by applying XDS scale factors used to similarly normalize the Bragg intensities. We assumed that large, radially symmetric changes in the measured scattering as the crystal is rotated originate from sources of non-crystalline scattering instead, such as solvent or coating oil. Based on this assumption, the largest-variance components between the radial intensity profiles were determined and removed from each image. This approach was validated by comparing symmetry-equivalent positions in the corrected maps, which were found to be consistent. This observation supports our initial assumption that large variance components were indeed caused by scattering from sources other than the protein crystal. 

In the case of CypA, a radially symmetric peak at $\vert \mathbf{q} \vert=1.3$ \AA$\mathrm{^{-1}}$ was variably observed, consistent with scattering from the paratone oil in which the crystal was coated prior to data collection. In images without visible paratone contamination, the experimental radial intensity profiles approximately followed a second-degree polynomial in the neighboring region of 0.63 $<$ $\vert \mathbf{q} \vert$ $<$ 1.88 \AA$^{-1}$. Based on this observation, each image's radial intensity profile was fit in this region of $\vert \mathbf{q} \vert$ to the sum of a second-degree polynomial and a scaled paratone profile: $a$$\vert \mathbf{q} \vert$$^2$ + $b$$\vert \mathbf{q} \vert$ + $c$ + $m$I$_\textrm{ref,paratone}$($\vert \mathbf{q} \vert$ - $q_0$). The reference paratone profile, I$_\textrm{ref,paratone}$($\vert \mathbf{q} \vert$), was derived from the computed structure factors for non-crystalline Paratone-N \cite{nonbragg}. Fitted parameters associated with the paratone profile were a multiplicative scale factor, $m$, and an offset in $\vert \mathbf{q} \vert$, $q_0$. The latter parameter accounts for anisotropy in the distribution of the paratone coating the crystal; optimized values of $q_0$ were small, with a mean of $q_0$ = 0.007 \AA$^{-1}$ across the 360 images in the CypA dataset. The resulting paratone profile was then subtracted from each image. Principal component analysis was performed to remove residual radial variance from the corrected images, and a background profile for each image was generated from the sum of the first two principal components, which were scaled by their associated eigenvalues and projected onto the image's radial intensity profile. The intensity correction for each pixel was then estimated by linear interpolation from these background profiles and subtracted from each image. In the case of WrpA, the first principal component was similarly used to remove variance in the radial intensity distributions across the rotation range. No further corrections were applied to the AP diffraction images. 

Diffuse scattering maps were constructed as three-dimensional grids in reciprocal space whose nodes oversample Miller indices by a factor of three along each lattice direction. Corrected intensities were binned into voxels centered on these nodes, and the mean pixel intensity of each voxel was used to estimate the intensity at each node. The signal-to-noise ratio was estimated as the mean divided by the standard deviation of the intensities binned into each voxel and is shown across resolution shells in Fig. S1. Maps were symmetrized by averaging the intensities of Laue- and Friedel-equivalent voxels, followed by subtraction of the interpolated radial intensity. If symmetrization is justified, this order of operations will provide a better estimate of the intensities used to compute the radial average profile; if not, it will introduce bias. However, reversing the symmetrization and radial average subtraction steps was not observed to affect the results, with all CCs within 0.01 of their prior values. A constant value was then added uniformly to each symmetrized, radial average-subtracted map to ensure that all intensities were positive. Code used to generate the maps is available at https://github.com/apeck12/diffuse.

\subsection{Bragg data processing} Refined structural models from the analysis of the Bragg component of these datasets have previously been reported  \cite{pmid27035972,pmid27189921,pmid26858101}. However, because the diffuse maps were generated using refined parameters and a modified version of the spot prediction algorithm from XDS, the Bragg data were reprocessed to maximize consistency between the treatment of the Bragg and diffuse signal. The Bragg data were indexed, integrated, and scaled with XDS \cite{pmid20124692}; statistics are shown in Table 1. For CypA, molecular replacement was performed with Phaser \cite{pmid19461840} using PDB 2CPL as a search model. This was followed by five macrocycles of refinement in Phenix \cite{pmid20124702} as previously described \cite{pmid27035972}. For AP, molecular replacement was performed with Phaser using wild-type AP (PDB 3TG0) stripped of non-protein atoms as the search model. As in Ref. \cite{pmid27189921}, zinc ions at full occupancy and a tungstate ion and water molecules at partial occupancy were manually modeled into the residual electron density in each active site. Automated refinement was performed using REFMAC5 \cite{pmid21460454}. For WrpA, molecular replacement was performed with Phaser using PDB 5F51 as a search model \cite{pmid26858101}. This was followed by alternating rounds of manual refinement in Coot \cite{pmid20383002} to model a sulfate ion and water molecules and automated refinement in Phenix. The R$_\textrm{work}$/R$_\textrm{free}$ values of the final refined models were similar to those previously deposited \cite{pmid27035972,pmid27189921,pmid26858101}. Diffraction images for the CypA, AP, and WrpA datasets are available at the SBGrid Data Bank, with accession numbers 68 \cite{sbdb68}, 456 \cite{sbdb456}, and 203 \cite{sbdb203}, respectively.

\subsection{Diffuse scattering predictions from real-space models of disorder} Experimental diffuse scattering maps were compared to the following disorder models:
\begin{itemize}
\item Gaussian elastic network model, a commonly used normal-mode decomposition of the protein motions based on the structure. Normal modes were computed using a uniform spring constant for all atom pairs within a given distance \cite{pmid22208195}, and the predicted interatomic correlations were renormalized by the B factors from the refined models of the Bragg data. 
\item Conformational ensemble models, which model configurational disorder as a discrete set of probability-weighted states \cite{guinier}. Conformational states were inferred by analyzing the crystal electron density from the Bragg data.
\item Rigid body rotations, in which the atoms in an asymmetric unit rotate as a unit around a random, isotropically oriented axis, with a normally distributed rotation angle \cite{pmid19836331}.
\item Rigid body translations, in which the atoms in an asymmetric unit translate as a unit. Translations sample an isotropic Gaussian distribution \cite{pmid19836331, pmid26863980}.
\item Liquid-like motions, in which correlations between atoms decay exponentially as a function of interatomic distance \cite{pmid9438860, pmid1603804}. Two forms of this model were considered: a variant in which correlations were confined within the boundaries of the asymmetric unit, and the conventional form of this model in which correlations extend between neighboring protein molecules, thereby crossing asymmetric unit and unit cell boundaries.
\end{itemize}
For all models except for the conventional liquid-like motions model, correlations were assumed to be confined within the boundaries of the asymmetric unit, with no coherence between neighboring molecules in the crystal lattice. For the systems considered here, the chosen asymmetric unit contained a single copy of the biological unit.

Diffuse scattering maps were simulated using Thor \cite{thor}, a software package for simulating and analyzing x-ray scattering experiments. For consistency with the experimental maps, the average radial intensity was subtracted from the predicted maps. Best fit parameters for each model were determined by scanning over the disorder parameter(s) to maximize CC with the experimental map. Agreement was assessed by the CC between the predicted and experimental maps, with each voxel downweighted by its multiplicity. For visual comparisons, a multiplicative scale factor and constant platform were applied to place the predicted maps on the same intensity scale as the experimental maps unless otherwise noted. Disorder models are described in more mathematical detail in the Supplementary Information.

\section{Results}

\begin{figure*}
\includegraphics[width=.8\linewidth]{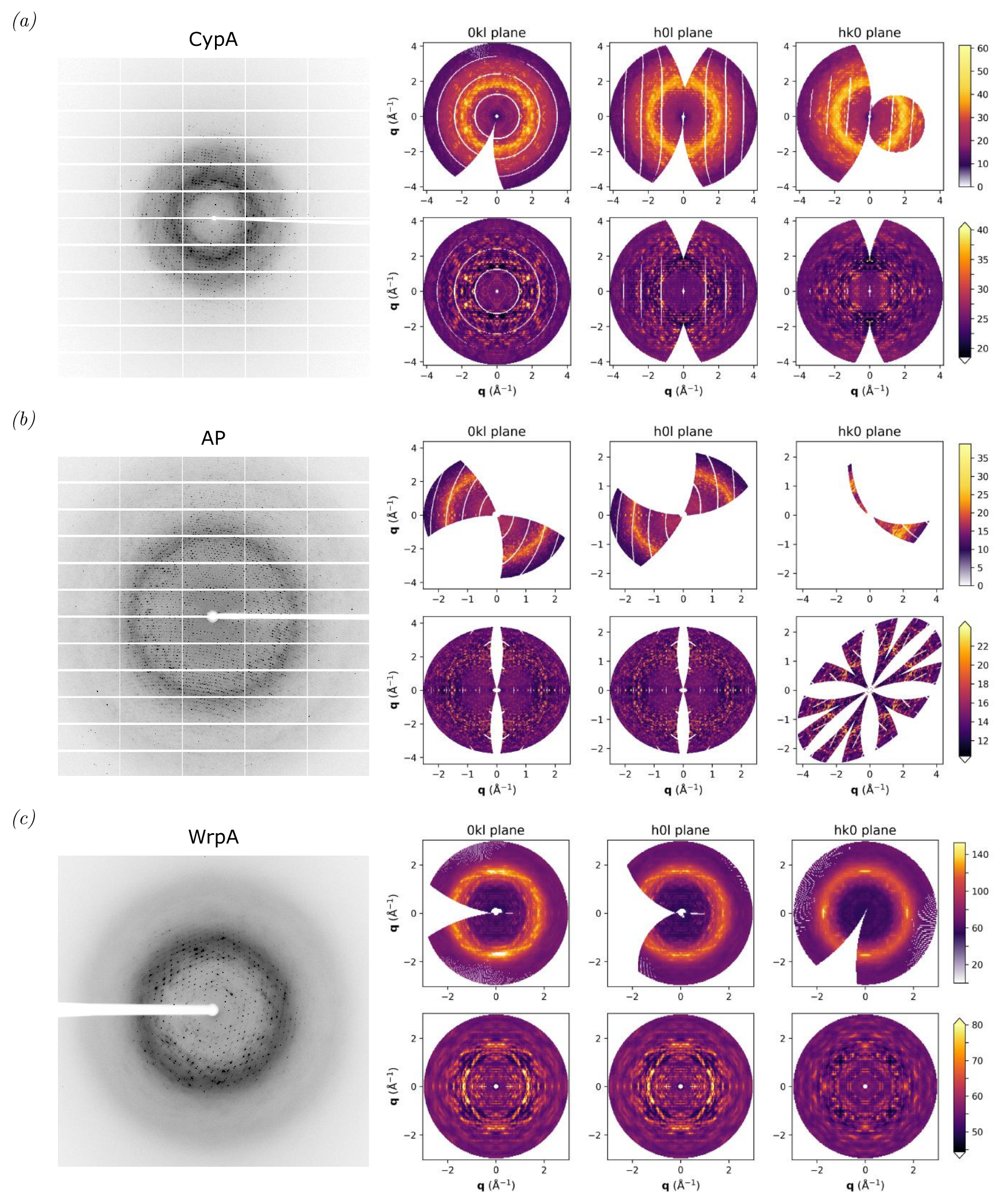}
\caption{\label{fig:experimental}\textbf{Reciprocal space maps from experimental diffuse scattering.} The diffuse scattering in diffraction images collected for \textit{(a)} CypA, \textit{(b)} AP, and \textit{(c)} WrpA was reconstructed into a reciprocal space map for each system. These maps take the form of three-dimensional grids that are 3x oversampled relative to the Miller indices along each lattice axis. \textit{(Left)} An example diffraction image from each dataset. \textit{(Right)} Central slices through reciprocal space are visualized for each unsymmetrized map in the top panels. The lower panels show these slices after symmetrization of Friedel- and Laue-equivalent voxels, followed by subtraction of the average radial intensity profile to highlight anisotropic features. For the symmetrized maps, the color scales do not span the entire range of voxel intensities; this saturates a subset of voxels but improves overall contrast.}
\end{figure*}

\subsection{The experimental maps exhibit Laue symmetry and significant anisotropic features}

We analyzed three crystallographic datasets collected by the rotation method for which diffuse scattering was visible in the raw diffraction images (Fig. 1, left). The Bragg data were separated and processed by standard protocols, yielding refined structural models similar to those previously published \cite{pmid27035972,pmid27189921,pmid26858101}. The diffuse scattering was isolated and processed to generate three-dimensional maps in reciprocal space (Fig. 1, upper right panels). The maps oversample the diffuse scattering signal along each lattice direction by a factor of three relative to the Miller indices, which enables these maps to resolve correlations that extend across multiple unit cells.

Overall statistics for the diffuse scattering maps are shown in Table~\ref{tab:stats} and by resolution shell in Fig. S1. The intensities of voxels related by Friedel's law and Laue symmetry showed significant correlation in all cases, supporting symmetrization of the maps by averaging the intensities of these symmetry-equivalent voxels. Currently there is no established convention for determining when the symmetrization of diffuse maps is justified, so we considered the CC between symmetry-related voxels to be significant based on the threshold value of CC$_{1/2}$ used to determine resolution cut-offs for the Bragg data \cite{pmid22628654}. Although these cases are not precisely analogous, this enhances the consistency between Bragg and diffuse data processing techniques. To remove the intensity contributions from uncorrelated disorder, which includes solvent and air scattering in addition to uncorrelated protein disorder, the average radial intensity was subtracted from each map. The resulting maps were characterized by significant anisotropic features, indicative of correlated disorder (Fig. 1, lower right panels).

\begin{figure}
\includegraphics{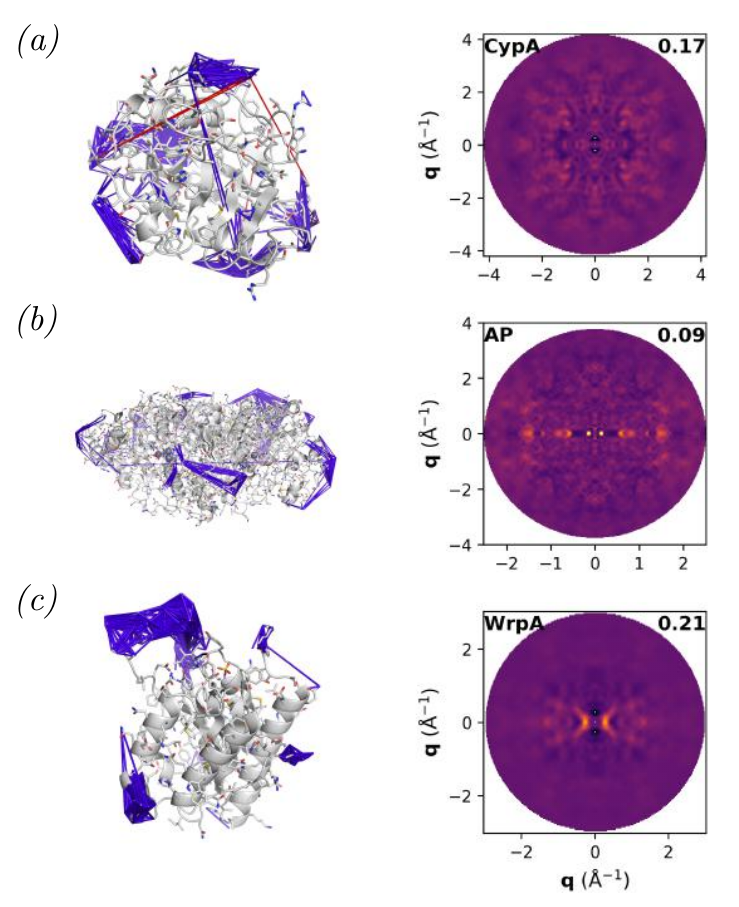}
\caption{\label{fig:network} \textbf{Elastic network models of Gaussian disorder.} The highest and lowest magnitude entries in the covariance matrix are overlaid as blue and red cylinders, respectively, on the refined atomic coordinates for \textit{(a)} CypA, \textit{(b)} AP, and \textit{(c)} WrpA. The 0kl slices of the predicted diffuse scattering maps are shown on the right, with the overall CC noted in white. The color scales differ from Fig. 1 to enhance visualization of the features.}
\end{figure}

\begin{figure}
\includegraphics{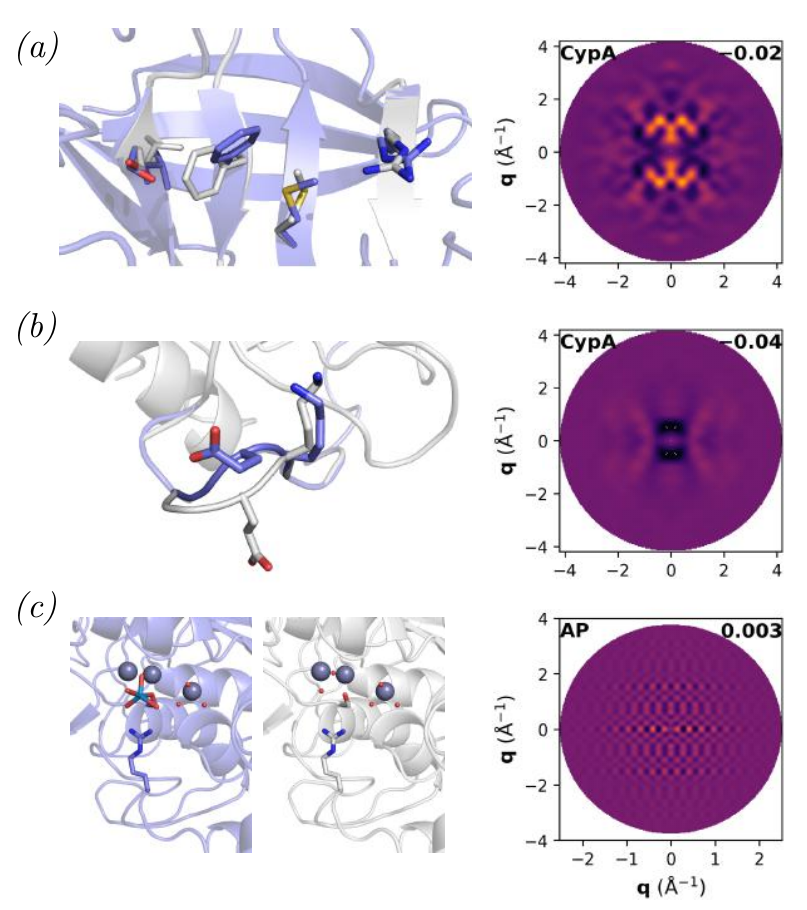}
\caption{\label{fig:ensemble} \textbf{Ensemble models inferred from the Bragg data.} \textit{(a)} Multi-conformer modeling of the CypA electron density map predicts a minor population of alternate conformers (purple) that radiate from the active site. \textit{(b)} CypA datasets collected at 180 K and below show two loop conformations between residues 79 and 83; the visualized loop conformations are those modeled in the 100 K dataset. Above 180 K, the conformation shown in white is not populated. \textit{(c)} Model of AP assuming that one active site is tungstate-bound (left) while the other is occupied by water molecules that coordinate the active site metal ions (right). The 0kl plane of the predicted map from each ensemble model is shown on the right, with a different color scale from Fig. 1 to enhance visual contrast. The overall CC between the predicted and experimental maps is noted in white.}
\end{figure}

\subsection{Models of conformational dynamics do not correlate with the experimental maps}

Models of intramolecular disorder that predict idiosyncratic configurational dynamics -- the type of motions most likely related to biological function -- were assessed for their ability to reproduce the experimental signal. A general class of these models assumes that interatomic displacements are small and sample a Gaussian distribution, and can thus be described by a covariance matrix. Here, covariance matrices were predicted from normal modes analysis of each protein structure in torsion angle space, using a standard form of the elastic network model that has been validated against Bragg-derived crystal structures \cite{pmid22208195}. Interatomic covariances were renormalized by the refined B factors such that the predicted amplitudes of motion were consistent with the structural models of the Bragg data. These elastic network models predicted distributions of strongly co-varying atom pairs that were non-uniform and often spatially localized in the protein (Fig. 2, left and Fig. S3). For all three systems, the diffuse scattering predicted by these network models was unable to reproduce the observed signal, apparent both in the low linear correlation coefficient (CC) and by visual comparison of the predicted and experimental 0kl planes (Fig. 2). 

Non-Gaussian ensemble models inferred from the Bragg data were also evaluated to determine whether such conformational heterogeneity contributes measurably to the diffuse signal. In the case of CypA, multiconformer modeling of the electron density map revealed a minor population of alternative rotamers for a series of residues that radiate from the active site (Fig. 3a, left) \cite{pmid19770508}. This observation of a correlated rotameric switch is consistent with prior analysis of CypA crystal structures \cite{pmid19770508, pmid19956261, pmid21918110, pmid26422513}. Another ensemble was generated from the loop conformations populated by residues 79-83 in CypA crystals collected at or below 180 K (Fig. 3b, left) \cite{pmid26422513}. Although only one of these conformations is populated in the dataset analyzed here (which was collected at 273 K), this model offers a distinct example of a type of configurational disorder prevalent in proteins. A third ensemble model was suggested by the occupancy disorder observed in AP, for which the Bragg coordinates were refined with a half-occupied tungstate ion in each active site. However, the Bragg data cannot distinguish between this model of partial occupancy and a model of correlated occupancy in which only one AP monomer is tungstate-bound at a given time (Fig. 3c, left). Whereas partial occupancy contributes to radially symmetric diffuse scattering, correlated occupancy yields anisotropic features. 

Diffuse scattering maps were predicted for each two-state model by Guinier's equation \cite{guinier}; the 0kl planes are shown in Fig. 3. The predicted maps from the CypA ensemble models are distinct from one another, but both exhibit features spread over much broader regions of reciprocal space (due to the short length scale of the disorder in real space) than observed experimentally (Figs. 1a vs. 3a-b). The map predicted by the correlated occupancy model shows a unique checkered pattern (Fig. 3c), but these regular features are similarly larger than the features observed in the experimental map for AP (Fig. 1c). Though diffuse scattering has been suggested as a route for validating conformational heterogeneity modeled during Bragg refinement \cite{pmid19836331, pmid23985728}, these ensemble models do not appreciably account for the diffuse signal in these datasets (Table~\ref{tab:cc}).

\squeezetable 
\begingroup
\begin{table}
\caption{\label{tab:cc}Correlation coefficients between predicted and experimental maps}
\begin{ruledtabular}
\begin{tabular}{cccccc}
& \multicolumn{5}{c}{Model} \\ \hline
& Rigid body & Rigid body & Liquid-like$\footnote{Values before and after the vertical bar indicate the asymmetric unit-confined and ``with neighbors'' models, respectively.}$ & Elastic & Ensemble \Tstrut \\
& rotations & translations &  motions & network & \\ \hline
CypA & 0.46 & 0.44 & 0.48 $\mid$ 0.71 & 0.17 & -0.02, -0.04$\footnote{Rotamer switch and disordered loop models, respectively.}$ \Tstrut \\
AP & 0.26 & 0.32 & 0.32 $\mid$ 0.66 & 0.09 & 0.00$\footnote{Correlated occupancy model.}$ \\
WrpA & 0.44 & 0.41 & 0.48 $\mid$ 0.67 & 0.21 & -- \\
\end{tabular}
\end{ruledtabular}
\end{table}
\endgroup

\subsection{Short-range rigid body and liquid-like motions models show modest agreement with the observed signal}

\squeezetable 
\begingroup
\begin{table}
\caption{\label{tab:params}Refined model parameters}
\begin{ruledtabular}
\begin{tabular}{cccccc}
& \multicolumn{5}{c}{Model} \\ \hline
& Bragg$\footnote{The Bragg $\sigma$ was computed from the Wilson B factor.}$ & Rigid body & \multicolumn{2}{c}{Liquid-like} & Rigid body \Tstrut \\
& Wilson B & translations & \multicolumn{2}{c}{motions$\footnote{The values to the left and right of the vertical bar indicate the asymmetric unit-confined model and conventional model `with neighbors,' respectively.}$} & rotations \\
& $\sigma$ (\AA) &  $\sigma$ (\AA) & $\sigma$ (\AA) & $\gamma$ (\AA) \footnote{The correlation length affects the volume in reciprocal space across which intensities of the crystal transform are blurred. Consequently, this parameter may be sensitive to how finely the diffuse signal is sampled. However, fitting the LLM to experimental maps constructed with 5x-oversampling relative to the Miller indices yielded similar values of $\gamma$, suggesting that the experimental features are sufficiently resolved with 3x-oversampling.} & $\sigma$ ($^{\circ}$) \\ \hline
CypA & 0.45 & 0.68 & 0.36 $\mid$ 0.39$\footnote{Prior analysis of this dataset found best fit parameters of $\sigma=$ 0.38 \AA{} and $\gamma=$ 7.1 \AA{} for the model with neighbors \citep{pmid27035972}. However, in that study the diffuse signal was sampled at integral Miller indices, and diffuse halos around Bragg peaks were suppressed. }$ & 18 $\mid$ 18$^\text{c}$ & 2.9 \Tstrut \\ 
AP & 0.63 & 0.63 & 0.40 $\mid$ 0.48 & 118 $\mid$ 53 & 0.9 \\
WrpA & 0.90 & 1.05 & 0.54 $\mid$ 0.61 & 15 $\mid$ 18 & 3.4 \\
\end{tabular}
\end{ruledtabular}
\end{table}
\endgroup

\begin{figure*}
\includegraphics[scale=0.9]{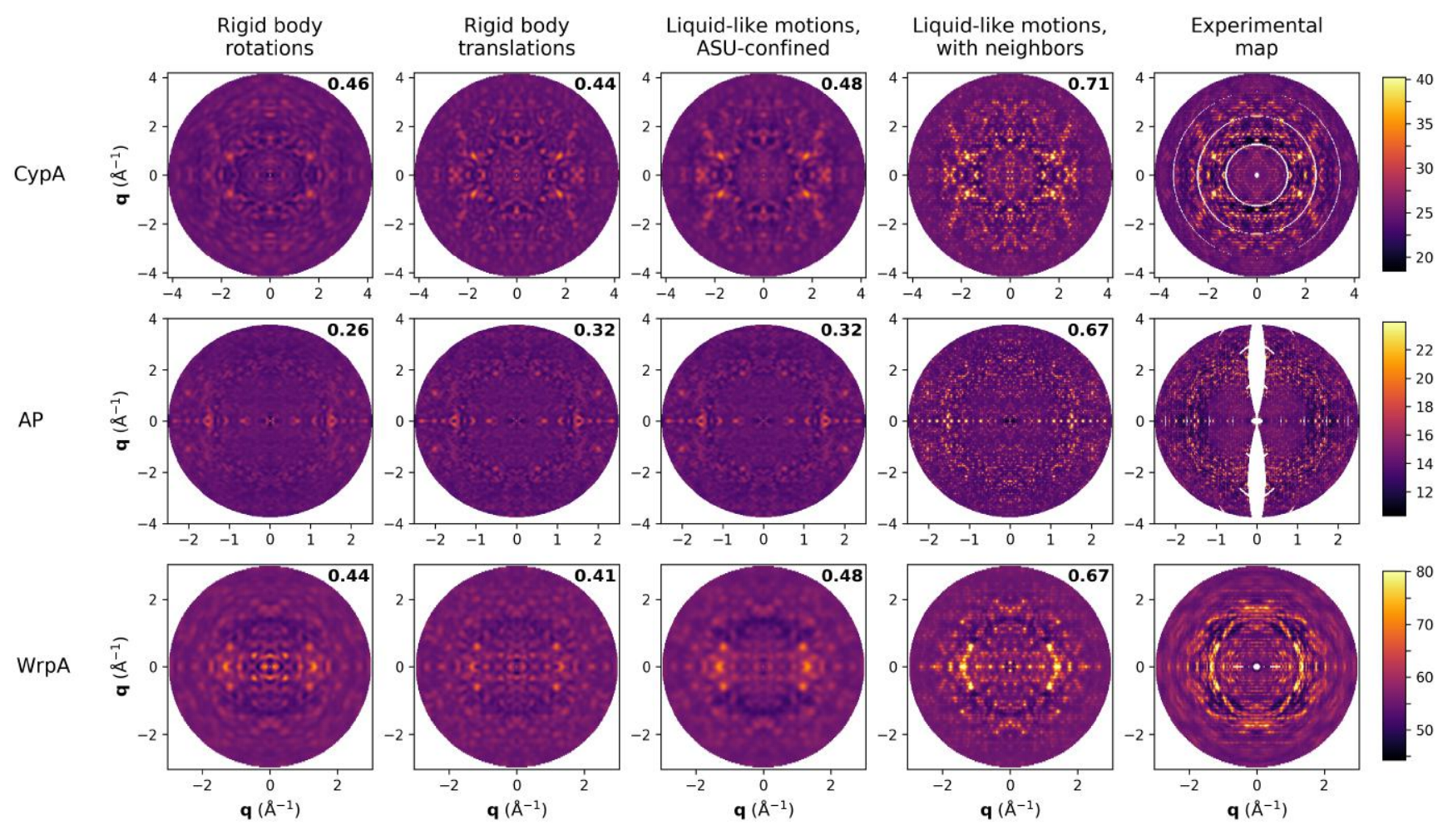}
\caption{\label{fig:simple}\textbf{Comparison of models of rigid body and liquid-like motions.} For the indicated model, parameters that tune the disorder were fit either by a grid scan or least-squares optimization to maximize overall correlation with the experimental map. The predicted 0kl planes for the best fit maps are shown, with the experimental 0kl planes displayed in the rightmost column for comparison. Overall correlation coefficients between the experimental and predicted maps are noted in white.}
\end{figure*}

The inability of these elastic network and ensemble models to reproduce the experimental maps prompted us to evaluate other disorder models which predict simpler dynamics of rigid body or liquid-like motions. As in the prior section, the models described below assume that correlations do not extend beyond the boundaries of individual asymmetric units. Additionally, the models evaluated in this section share the symmetrized molecular transform -- specifically, the Fourier transform of the individual protein molecule, incoherently summed over its orientations in the unit cell -- as their basis, which, as discussed below, raises the possibility of using the diffuse scattering signal for static structural inference.

The diffuse scattering predicted by rigid body rotational disorder is related to the variance of an ensemble of rotated structure factors. Visually, this type of disorder has the effect of blurring features of the molecular transform in concentric shells of reciprocal space. An isotropic version of this model showed modest correlation with the CypA and WrpA maps (Fig. 4). For both maps, the best fit values for the standard deviation of the angle of rotation were on the order of 2-3$^{\circ}$ (Table~\ref{tab:params}), consistent with a blurring effect that spans a few voxels of these reciprocal space maps. The best fit value for the AP map was smaller (0.9$^{\circ}$), yielding minimal radial blurring that could be resolved by the coarseness of the map's voxels, which along with the modest correlation suggested that rotational disorder was inconsistent with the observed signal. Relative to CypA and WrpA, AP has more crystal contacts that may inhibit this type of disorder. It is also possible that the finer slicing during collection of the AP data minimized blurring, but radial blurring due to data collection versus as a result of rotational disorder in the crystal cannot be distinguished by the isotropic model considered here.

By contrast, the diffuse scattering produced by rigid body translational disorder is the molecular transform scaled by the Debye-Waller factor \cite{pmid19836331,pmid26863980}. For all three maps, this disorder model showed nontrivial correlation with the experimental maps. Further, the best fit values of the isotropic displacement parameter $\sigma$, which reports on the scale of displacement, were within two-fold of the value predicted by the Bragg Wilson B factor (Fig. 4, Table~\ref{tab:params}), suggesting that the diffuse signal is consistent with scattered intensity missing in the Bragg data due to disorder. The fit was modestly improved by imposing exponential decay on interatomic covariances, thereby switching from a rigid body to a liquid-like description of correlated dynamics (Fig. 4). This model of asymmetric unit-confined liquid-like motions predicted similar though consistently smaller values for the isotropic displacement parameter, and in the case of CypA and WrpA, a correlation length roughly one third to one half the dimensions of the protein molecule (Table~\ref{tab:params}). In the case of AP, the best fit correlation length spanned the longest dimension of the protein, consistent with the lack of improvement in CC: in the regime of correlation lengths longer than the protein unit, the diffuse scattering predictions of the liquid-like motions and rigid body translational disorder models converge. 

\subsection{Speckles indicate long-range correlated disorder that crosses unit cell boundaries}

\begin{figure*}
\includegraphics[width=.9\linewidth]{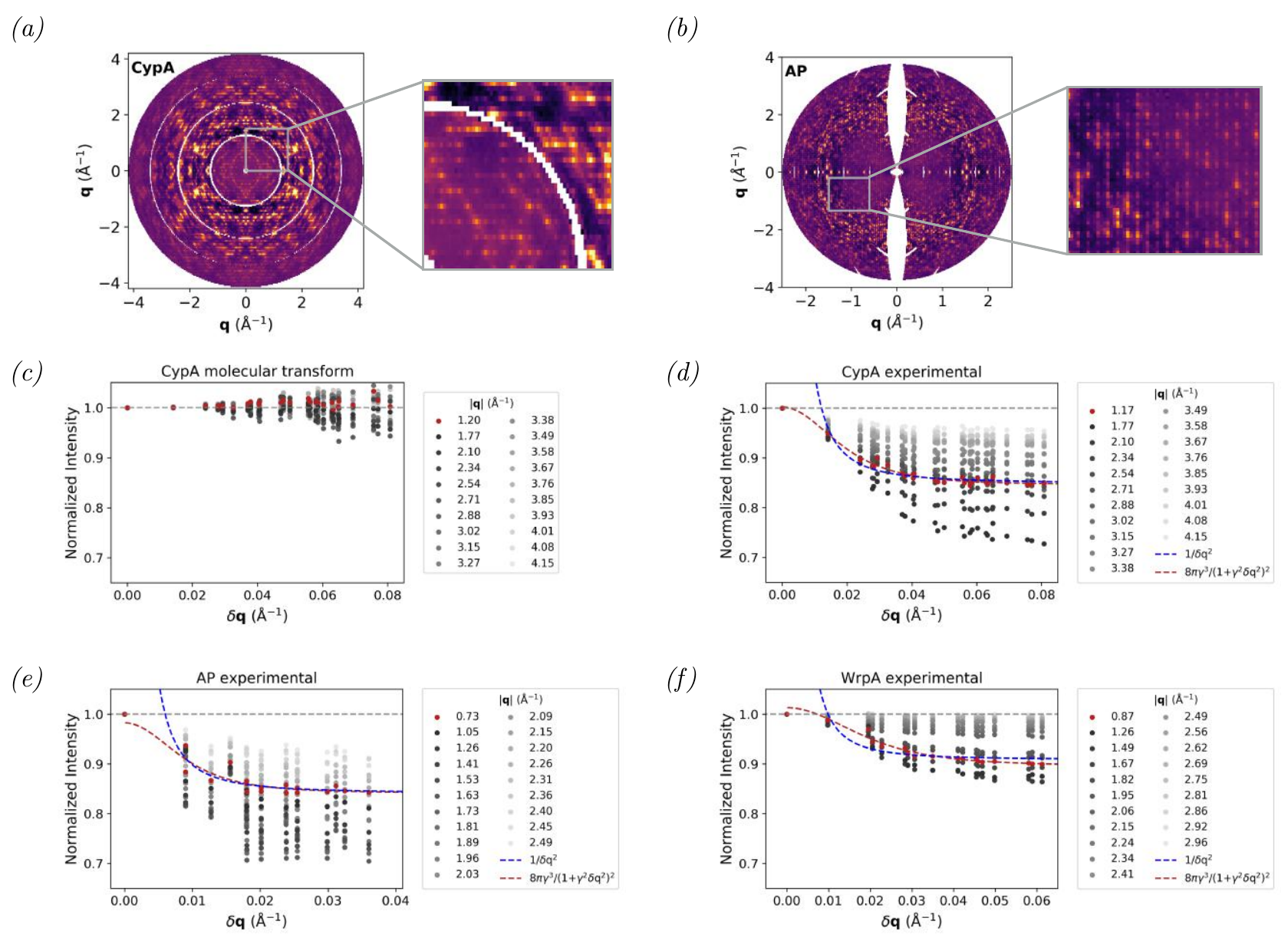}
\caption{\label{fig:phonons}\textbf{Speckled features of the experimental maps.} Insets of the 0kl planes of the \textit{(a)} CypA and \textit{(b)} AP maps highlight the characteristic speckles. For \textit{(c-f)}, maps were reconstructed or computed with 5x oversampling along each lattice direction relative to integral Miller indices. Data were binned into twenty resolution shells, and the median profile for the intensity as a function of $\delta\mathbf{q}$, the distance from reciprocal lattice sites, is shown for each shell. Intensity profiles were normalized such that the intensity value for voxels that coincided with reciprocal lattice sites ($\delta\mathbf{q}$ = 0) was unity. For each experimental map, curves with a $1/\delta\mathbf{q}^2$ (dashed blue) or a $8\pi\gamma^3/(1+\gamma^2\delta\mathbf{q}^2)^2$ (dashed red) dependence, as predicted by the phonon and liquid-like motions models, respectively, were fit to the intensity profile in the lowest resolution shell (red points). For comparison to models whose basis is the molecular transform, intensity profiles for the CypA molecular transform map are shown in \textit{(c)}. }
\end{figure*}

Of the models considered above, in no case does the correlation coefficient between the predicted and experimental map exceed 0.5 (Table ~\ref{tab:cc}, Fig. S4). Visual inspection suggests that a feature these models systematically fail to reproduce is the observed ``speckles,'' periodic spikes in intensity that appear superimposed on diffuse scattering features that span larger volumes in reciprocal space (Fig. 5a-b, insets). Such speckles arise from enhanced scattering at reciprocal lattice positions -- i.e., the estimated diffuse intensity underneath Bragg peaks -- and have previously been noted in studies that analyzed the diffuse scattering signal at fractional Miller indices \cite{bw0632, pmid26457426, pmid17930640}. Because the length scale of disorder in real space determines the spacing of diffuse features in reciprocal space, the need to oversample the diffuse signal relative to integral Miller indices to observe these speckles indicates that they arise from correlations that extend beyond the boundaries of a single unit cell. The models examined in prior sections assumed correlated disorder confined within asymmetric units, so were unable to generate this type of signal.

We therefore considered two models of disorder in which correlations extend across unit cell boundaries to determine if accounting for intermolecular correlations improved predictions of the total diffuse signal, including these speckled features. The first of these models is the traditional liquid-like motions model, in which the basis is the crystal rather than the molecular transform. For each system, this model of long-range liquid-like motions showed considerable agreement with the experimental signal, qualitatively reproducing the speckled features and quantitatively yielding the highest CC of the models considered here (Fig. 4, Table ~\ref{tab:cc}). This improved correlation is observed not just at the reciprocal lattice sites where the speckles are centered, but also at map voxels farthest from integral Miller indices (Fig. S4b). The refined model parameters for the two liquid-like motions models were similar in most cases (Table ~\ref{tab:params}), indicating that the consistent increase in CC between the short- and long-range models resulted almost exclusively from taking into account correlations across neighbors in the crystal, which contribute to the diffuse signal throughout reciprocal space. Further support for this liquid-like motions model comes from a comparison of the predicted and experimental autocorrelation functions, from which correlation lengths in real space can be inferred. Peaks consistent with the unit cell dimensions, and thus indicative of correlations extending across unit cell boundaries, were observed in the autocorrelation function of each experimental map and the long-range liquid-like motions model. By contrast, models in which correlations were confined within the boundaries of the asymmetric unit did not reproduce these characteristic peaks (Fig. S5).

An alternative model proposed to account for speckled features invokes acoustic lattice vibrations from phonon-induced inelastic scattering \cite{bw0632, pmid26457426, pmid17930640}. One prediction of this model is that the diffuse intensity will decrease proportional to the square of the distance from reciprocal lattice positions, $\delta\mathbf{q}$. Such a trend is absent from the molecular transform and its derivative models (Fig. 5c), in which disorder is confined within the boundaries of asymmetric units. Though qualitatively the phonon model accounts for the observed halos around Bragg peaks, quantitatively the experimental fall-off in intensity differs from the $1/\delta\mathbf{q}^2$ dependence predicted for single phonon interactions and is better fit by the dependence predicted by the liquid-like motions kernel (Fig. 5d-f, dashed blue versus red) \cite{bw0632, pmid17930640}. More complex phonon models, from either extending the spectrum to include optical modes or accounting multiple-phonon effects, are predicted to cause intensity to vary more slowly with distance from reciprocal lattice sites \cite{bw0632}. However, there is currently no robust method for predicting the diffuse scattering produced by phonons in macromolecular crystals. In the absence of such a method and an established procedure for simulating competing acoustic modes, let alone optical modes or the effects of multiple phonon interactions, we cannot fully assess agreement with the phonon model versus other types of long-range disorder.

\section{Discussion}

Here we present a unified framework of the principal disorder models that have previously been used to interpret diffuse scattering, and compare their ability to reproduce the signal observed in three experimental datasets. Consistent with prior work, the above analysis finds that rigid body and liquid-like motions models exhibit modest correlation with the experimental maps when correlated disorder is confined to the asymmetric unit \cite{pmid26863980}. Multiple models that predict more complex intramolecular dynamics were also considered, but showed minimal agreement with experiment (Table ~\ref{tab:cc}). Experimentally-observed speckles did not fit the profile for phonon-induced lattice dynamics, but in agreement with prior results, could largely be reproduced by the conventional form of the liquid-like motions model, in which disorder extends across neighboring asymmetric units and thus unit cell boundaries \cite{pmid9438860, pmid3808065, pmid27035972}. None of the models assessed here fully explained the experimental signal in these datasets, which represent a range of crystallographic properties and biological functions. However, all three protein systems were globular proteins, and it is possible that distinct types of disorder underlie the diffuse scattering from crystals of membrane and fibrous proteins. 

Past interest in diffuse scattering has primarily stemmed from the premise that these data probe dynamics related to biological function \cite{pmid24507780, pmid9438860, pmid9636718, pmid15299461, pmid15681654, pmid1600083, pmid16384188, pmid17154425, pmid25453071, pmid27035972}. However, the experimental maps showed minimal correlation with the elastic network and ensemble models assessed here. These specific models represent a limited subspace of possible models of conformational dynamics that are consistent with the Bragg data, and it is likely that refining the models' parameters could improve agreement with the diffuse signal. However, the observation of experimental features indicative of correlations that span neighboring molecules in the crystal cautions against the assumption that the dominant signal originates from the same protein motions that occur under physiological conditions, which these models attempt to capture. Models of disorder that account for both intermolecular and intramolecular correlations will thus be needed to resolve the contributions of each to the observed signal, a prerequisite in determining whether diffuse scattering is a useful method for studying dynamics associated with biological function.

On the other hand, the ability of rigid body and liquid-like motions models to reproduce many experimental features suggests that diffuse scattering data could in some cases be useful for resolution extension or phase retrieval. These models share the molecular or crystal transform as their basis, and thus yield a scaled or blurred image of this transform in the diffuse scattering map. The general observation that the diffuse scattering does not \emph{directly} reflect the molecular or crystal transform, but at the very least the convolution of the transform with some blurring function, must be better understood and taken into account.

Extracting the molecular transform signal will be particularly challenging for maps that exhibit enhanced scattering at reciprocal lattice sites, as observed here. The conventional liquid-like motions model, in which correlations are not confined to the asymmetric unit but rather extend between neighboring units in the crystal, best accounted for this feature. However, in the linear approximation, this model is a convolution of the disorder-free diffraction with a kernel that is the Fourier representation of the disorder. Thus, this model reports on the crystal transform, which is non-zero only at integral Miller indices. It does not contain information about the value of the molecular transform at fractional Miller indices. Iterative phase retrieval algorithms, such as those recently used by Ayyer \textit{et. al.} \cite{pmid26863980}, require these non-integral oversampled measurements to uniquely determine unmeasured phases \cite{Sayre:a00763}. Thus, the information present in the liquid-like motions model could in principle be employed for resolution extension, but not phase retrieval. However, the liquid-like motions model is approximate, and a more rigorous treatment of crystalline disorder may enable measurement of an oversampled molecular transform from the diffuse scattering. Despite this possibility, our results call into question the practice of directly using a diffuse scattering map for either resolution extension or iterative phasing in cases that exhibit enhanced scattering at reciprocal lattice positions, a feature observed in all three systems we studied, and one that we have no physical or theoretical grounds to mask or model separately from the remainder of the diffuse signal. Precisely how to deconvolve useful signals from such maps remains an open area of investigation.

The above analysis highlights a need for new models of diffuse scattering, either to interpret biologically relevant disorder or improve structure determination. If different sources of disorder are largely uncoupled, their contributions to the diffuse scattering will be approximately additive. However, the absence of coupling between distinct types of motions is not guaranteed, particularly in the context of a crystal lattice. Thus, challenges lie ahead both in jointly modeling distinct sources of disorder and in deconvolving weak signals from dominant features. The search space for such models is intractably large, so the number of acceptable free parameters and constraints will require careful treatment. Ideally, it would be possible to assert a model sophisticated enough to report interesting and idiosyncratic disorder in different systems (such as functional motions), but simple enough (\textit{i.e.} with few independent parameters) to infer directly from the observed data.

An alternative route is detailed forward modeling, such as molecular dynamics, which has previously been used to analyze diffuse scattering \cite{pmid7656016,pmid9636718,pmid25453071}. Molecular dynamics concurrently simulates multiple types of disorder, but this method does not lend itself to refining the contributions of different kinds of disorder to fit experimental data. In the common case where such simulations do not satisfactorily reproduce experimental observations, it is challenging to modify them in a principled manner so that they do. Combined with the computational expense of these methods, it seems prudent to seek simple explanations and models for analyzing diffuse scattering before comparing to atomic simulation. The incisive test of any model will come from its predictive power: confirming that a specific physical perturbation of a crystal system results in the predicted change to the diffuse signal.

\section{Conclusions}
Here we investigated the physical origins of the diffuse scattering observed from three protein crystals. A comprehensive comparison of previously proposed models critically addressed the nature and length scale of the disorder underlying this signal. Multiple models of intramolecular conformational dynamics, including ensemble models inferred from the Bragg data, were unable to explain the observed diffuse scattering. Whereas models of rigid body and liquid-like motions of individual proteins consistently showed modest agreement with experiment, a model of extended liquid-like motions across the crystal achieved high correlations with the three datasets analyzed (CCs $\sim$ 0.7). This analysis indicates that accounting for the intermolecular component of the disorder will be critical to successfully model this signal, which, in turn, is necessary to interpret diffuse scattering either to probe conformational dynamics or to enhance static structure inference.

\begin{acknowledgments}
We thank Vijay Pande, Soichi Wakatsuki, and Peter Moore for instructive discussions, and Dan Herschlag for the alkaline phosphatase dataset. We also thank Dan Herschlag, Nozomi Ando, Michael Wall, James Fraser, Henry Chapman, and Kartik Ayyer for comments on the manuscript. Michael Levitt is gratefully acknowledged for providing his Normal Mode Analysis code. The Stanford Research Computing Center provided the Sherlock cluster and other computational resources that facilitated this work. A.P. was funded by a National Science Foundation Graduate Research Fellowship and supported by Vijay Pande. F.P. acknowledges support from the National Institutes of Health (NIH), grant No. R35GM122543 (PI: Michael Levitt). T.J.L. was supported by the LCLS directorate, SLAC National Accelerator Laboratory, via the U.S. Department of Energy Office of Basic Energy Sciences Contract No. DE-AC02-76SF00515.
\end{acknowledgments}

\bibliography{apssamp}

\providecommand{\noopsort}[1]{}\providecommand{\singleletter}[1]{#1}
\begin{thebibliography}{47}%
\makeatletter
\providecommand \@ifxundefined [1]{%
 \@ifx{#1\undefined}
}%
\providecommand \@ifnum [1]{%
 \ifnum #1\expandafter \@firstoftwo
 \else \expandafter \@secondoftwo
 \fi
}%
\providecommand \@ifx [1]{%
 \ifx #1\expandafter \@firstoftwo
 \else \expandafter \@secondoftwo
 \fi
}%
\providecommand \natexlab [1]{#1}%
\providecommand \enquote  [1]{``#1''}%
\providecommand \bibnamefont  [1]{#1}%
\providecommand \bibfnamefont [1]{#1}%
\providecommand \citenamefont [1]{#1}%
\providecommand \href@noop [0]{\@secondoftwo}%
\providecommand \href [0]{\begingroup \@sanitize@url \@href}%
\providecommand \@href[1]{\@@startlink{#1}\@@href}%
\providecommand \@@href[1]{\endgroup#1\@@endlink}%
\providecommand \@sanitize@url [0]{\catcode `\\12\catcode `\$12\catcode
  `\&12\catcode `\#12\catcode `\^12\catcode `\_12\catcode `\%12\relax}%
\providecommand \@@startlink[1]{}%
\providecommand \@@endlink[0]{}%
\providecommand \url  [0]{\begingroup\@sanitize@url \@url }%
\providecommand \@url [1]{\endgroup\@href {#1}{\urlprefix }}%
\providecommand \urlprefix  [0]{URL }%
\providecommand \Eprint [0]{\href }%
\providecommand \doibase [0]{http://dx.doi.org/}%
\providecommand \selectlanguage [0]{\@gobble}%
\providecommand \bibinfo  [0]{\@secondoftwo}%
\providecommand \bibfield  [0]{\@secondoftwo}%
\providecommand \translation [1]{[#1]}%
\providecommand \BibitemOpen [0]{}%
\providecommand \bibitemStop [0]{}%
\providecommand \bibitemNoStop [0]{.\EOS\space}%
\providecommand \EOS [0]{\spacefactor3000\relax}%
\providecommand \BibitemShut  [1]{\csname bibitem#1\endcsname}%
\let\auto@bib@innerbib\@empty
\bibitem [{\citenamefont {Wall}\ \emph
  {et~al.}(2014{\natexlab{a}})\citenamefont {Wall}, \citenamefont {Adams},
  \citenamefont {Fraser},\ and\ \citenamefont {Sauter}}]{pmid24507780}%
  \BibitemOpen
  \bibfield  {author} {\bibinfo {author} {\bibfnamefont {M.~E.}\ \bibnamefont
  {Wall}}, \bibinfo {author} {\bibfnamefont {P.~D.}\ \bibnamefont {Adams}},
  \bibinfo {author} {\bibfnamefont {J.~S.}\ \bibnamefont {Fraser}}, \ and\
  \bibinfo {author} {\bibfnamefont {N.~K.}\ \bibnamefont {Sauter}},\
  }\href@noop {} {\bibfield  {journal} {\bibinfo  {journal} {Structure}\
  }\textbf {\bibinfo {volume} {22}},\ \bibinfo {pages} {182} (\bibinfo {year}
  {2014}{\natexlab{a}})}\BibitemShut {NoStop}%
\bibitem [{\citenamefont {Welberry}\ and\ \citenamefont
  {Weber}(2016)}]{welberry2016}%
  \BibitemOpen
  \bibfield  {author} {\bibinfo {author} {\bibfnamefont {T.}~\bibnamefont
  {Welberry}}\ and\ \bibinfo {author} {\bibfnamefont {T.}~\bibnamefont
  {Weber}},\ }\href {\doibase 10.1080/0889311X.2015.1046853} {\bibfield
  {journal} {\bibinfo  {journal} {Crystallography Reviews}\ }\textbf {\bibinfo
  {volume} {22}},\ \bibinfo {pages} {2} (\bibinfo {year} {2016})}\BibitemShut
  {NoStop}%
\bibitem [{\citenamefont {Moore}(2009)}]{pmid19836331}%
  \BibitemOpen
  \bibfield  {author} {\bibinfo {author} {\bibfnamefont {P.~B.}\ \bibnamefont
  {Moore}},\ }\href@noop {} {\bibfield  {journal} {\bibinfo  {journal}
  {Structure}\ }\textbf {\bibinfo {volume} {17}},\ \bibinfo {pages} {1307}
  (\bibinfo {year} {2009})}\BibitemShut {NoStop}%
\bibitem [{\citenamefont {Benoit}\ and\ \citenamefont
  {Doucet}(1995)}]{pmid7568674}%
  \BibitemOpen
  \bibfield  {author} {\bibinfo {author} {\bibfnamefont {J.~P.}\ \bibnamefont
  {Benoit}}\ and\ \bibinfo {author} {\bibfnamefont {J.}~\bibnamefont
  {Doucet}},\ }\href@noop {} {\bibfield  {journal} {\bibinfo  {journal} {Q.
  Rev. Biophys.}\ }\textbf {\bibinfo {volume} {28}},\ \bibinfo {pages} {131}
  (\bibinfo {year} {1995})}\BibitemShut {NoStop}%
\bibitem [{\citenamefont {Wall}\ \emph {et~al.}(1997)\citenamefont {Wall},
  \citenamefont {Clarage},\ and\ \citenamefont {Phillips}}]{pmid9438860}%
  \BibitemOpen
  \bibfield  {author} {\bibinfo {author} {\bibfnamefont {M.~E.}\ \bibnamefont
  {Wall}}, \bibinfo {author} {\bibfnamefont {J.~B.}\ \bibnamefont {Clarage}}, \
  and\ \bibinfo {author} {\bibfnamefont {G.~N.}\ \bibnamefont {Phillips}},\
  }\href@noop {} {\bibfield  {journal} {\bibinfo  {journal} {Structure}\
  }\textbf {\bibinfo {volume} {5}},\ \bibinfo {pages} {1599} (\bibinfo {year}
  {1997})}\BibitemShut {NoStop}%
\bibitem [{\citenamefont {Hery}\ \emph {et~al.}(1998)\citenamefont {Hery},
  \citenamefont {Genest},\ and\ \citenamefont {Smith}}]{pmid9636718}%
  \BibitemOpen
  \bibfield  {author} {\bibinfo {author} {\bibfnamefont {S.}~\bibnamefont
  {Hery}}, \bibinfo {author} {\bibfnamefont {D.}~\bibnamefont {Genest}}, \ and\
  \bibinfo {author} {\bibfnamefont {J.~C.}\ \bibnamefont {Smith}},\ }\href@noop
  {} {\bibfield  {journal} {\bibinfo  {journal} {J. Mol. Biol.}\ }\textbf
  {\bibinfo {volume} {279}},\ \bibinfo {pages} {303} (\bibinfo {year}
  {1998})}\BibitemShut {NoStop}%
\bibitem [{\citenamefont {Kolatkar}\ \emph {et~al.}(1994)\citenamefont
  {Kolatkar}, \citenamefont {Clarage},\ and\ \citenamefont
  {Phillips}}]{pmid15299461}%
  \BibitemOpen
  \bibfield  {author} {\bibinfo {author} {\bibfnamefont {A.~R.}\ \bibnamefont
  {Kolatkar}}, \bibinfo {author} {\bibfnamefont {J.~B.}\ \bibnamefont
  {Clarage}}, \ and\ \bibinfo {author} {\bibfnamefont {G.~N.}\ \bibnamefont
  {Phillips}},\ }\href@noop {} {\bibfield  {journal} {\bibinfo  {journal} {Acta
  Crystallogr. D Biol. Crystallogr.}\ }\textbf {\bibinfo {volume} {50}},\
  \bibinfo {pages} {210} (\bibinfo {year} {1994})}\BibitemShut {NoStop}%
\bibitem [{\citenamefont {Meinhold}\ and\ \citenamefont
  {Smith}(2005{\natexlab{a}})}]{pmid15681654}%
  \BibitemOpen
  \bibfield  {author} {\bibinfo {author} {\bibfnamefont {L.}~\bibnamefont
  {Meinhold}}\ and\ \bibinfo {author} {\bibfnamefont {J.~C.}\ \bibnamefont
  {Smith}},\ }\href@noop {} {\bibfield  {journal} {\bibinfo  {journal}
  {Biophys. J.}\ }\textbf {\bibinfo {volume} {88}},\ \bibinfo {pages} {2554}
  (\bibinfo {year} {2005}{\natexlab{a}})}\BibitemShut {NoStop}%
\bibitem [{\citenamefont {Chacko}\ and\ \citenamefont
  {Phillips}(1992)}]{pmid1600083}%
  \BibitemOpen
  \bibfield  {author} {\bibinfo {author} {\bibfnamefont {S.}~\bibnamefont
  {Chacko}}\ and\ \bibinfo {author} {\bibfnamefont {G.~N.}\ \bibnamefont
  {Phillips}},\ }\href@noop {} {\bibfield  {journal} {\bibinfo  {journal}
  {Biophys. J.}\ }\textbf {\bibinfo {volume} {61}},\ \bibinfo {pages} {1256}
  (\bibinfo {year} {1992})}\BibitemShut {NoStop}%
\bibitem [{\citenamefont {Meinhold}\ and\ \citenamefont
  {Smith}(2005{\natexlab{b}})}]{pmid16384188}%
  \BibitemOpen
  \bibfield  {author} {\bibinfo {author} {\bibfnamefont {L.}~\bibnamefont
  {Meinhold}}\ and\ \bibinfo {author} {\bibfnamefont {J.~C.}\ \bibnamefont
  {Smith}},\ }\href@noop {} {\bibfield  {journal} {\bibinfo  {journal} {Phys.
  Rev. Lett.}\ }\textbf {\bibinfo {volume} {95}},\ \bibinfo {pages} {218103}
  (\bibinfo {year} {2005}{\natexlab{b}})}\BibitemShut {NoStop}%
\bibitem [{\citenamefont {Meinhold}\ and\ \citenamefont
  {Smith}(2007)}]{pmid17154425}%
  \BibitemOpen
  \bibfield  {author} {\bibinfo {author} {\bibfnamefont {L.}~\bibnamefont
  {Meinhold}}\ and\ \bibinfo {author} {\bibfnamefont {J.~C.}\ \bibnamefont
  {Smith}},\ }\href@noop {} {\bibfield  {journal} {\bibinfo  {journal}
  {Proteins}\ }\textbf {\bibinfo {volume} {66}},\ \bibinfo {pages} {941}
  (\bibinfo {year} {2007})}\BibitemShut {NoStop}%
\bibitem [{\citenamefont {Wall}\ \emph
  {et~al.}(2014{\natexlab{b}})\citenamefont {Wall}, \citenamefont
  {Van~Benschoten}, \citenamefont {Sauter}, \citenamefont {Adams},
  \citenamefont {Fraser},\ and\ \citenamefont {Terwilliger}}]{pmid25453071}%
  \BibitemOpen
  \bibfield  {author} {\bibinfo {author} {\bibfnamefont {M.~E.}\ \bibnamefont
  {Wall}}, \bibinfo {author} {\bibfnamefont {A.~H.}\ \bibnamefont
  {Van~Benschoten}}, \bibinfo {author} {\bibfnamefont {N.~K.}\ \bibnamefont
  {Sauter}}, \bibinfo {author} {\bibfnamefont {P.~D.}\ \bibnamefont {Adams}},
  \bibinfo {author} {\bibfnamefont {J.~S.}\ \bibnamefont {Fraser}}, \ and\
  \bibinfo {author} {\bibfnamefont {T.~C.}\ \bibnamefont {Terwilliger}},\
  }\href@noop {} {\bibfield  {journal} {\bibinfo  {journal} {Proc. Natl. Acad.
  Sci. U.S.A.}\ }\textbf {\bibinfo {volume} {111}},\ \bibinfo {pages} {17887}
  (\bibinfo {year} {2014}{\natexlab{b}})}\BibitemShut {NoStop}%
\bibitem [{\citenamefont {Van~Benschoten}\ \emph {et~al.}(2016)\citenamefont
  {Van~Benschoten}, \citenamefont {Liu}, \citenamefont {Gonzalez},
  \citenamefont {Brewster}, \citenamefont {Sauter}, \citenamefont {Fraser},\
  and\ \citenamefont {Wall}}]{pmid27035972}%
  \BibitemOpen
  \bibfield  {author} {\bibinfo {author} {\bibfnamefont {A.~H.}\ \bibnamefont
  {Van~Benschoten}}, \bibinfo {author} {\bibfnamefont {L.}~\bibnamefont {Liu}},
  \bibinfo {author} {\bibfnamefont {A.}~\bibnamefont {Gonzalez}}, \bibinfo
  {author} {\bibfnamefont {A.~S.}\ \bibnamefont {Brewster}}, \bibinfo {author}
  {\bibfnamefont {N.~K.}\ \bibnamefont {Sauter}}, \bibinfo {author}
  {\bibfnamefont {J.~S.}\ \bibnamefont {Fraser}}, \ and\ \bibinfo {author}
  {\bibfnamefont {M.~E.}\ \bibnamefont {Wall}},\ }\href@noop {} {\bibfield
  {journal} {\bibinfo  {journal} {Proc. Natl. Acad. Sci. U.S.A.}\ }\textbf
  {\bibinfo {volume} {113}},\ \bibinfo {pages} {4069} (\bibinfo {year}
  {2016})}\BibitemShut {NoStop}%
\bibitem [{\citenamefont {Ayyer}\ \emph {et~al.}(2016)\citenamefont {Ayyer},
  \citenamefont {Yefanov}, \citenamefont {Oberthur}, \citenamefont
  {Roy-Chowdhury}, \citenamefont {Galli}, \citenamefont {Mariani},
  \citenamefont {Basu}, \citenamefont {Coe}, \citenamefont {Conrad},
  \citenamefont {Fromme}, \citenamefont {Schaffer}, \citenamefont {Dorner},
  \citenamefont {James}, \citenamefont {Kupitz}, \citenamefont {Metz},
  \citenamefont {Nelson}, \citenamefont {Xavier}, \citenamefont {Beyerlein},
  \citenamefont {Schmidt}, \citenamefont {Sarrou}, \citenamefont {Spence},
  \citenamefont {Weierstall}, \citenamefont {White}, \citenamefont {Yang},
  \citenamefont {Zhao}, \citenamefont {Liang}, \citenamefont {Aquila},
  \citenamefont {Hunter}, \citenamefont {Robinson}, \citenamefont {Koglin},
  \citenamefont {Boutet}, \citenamefont {Fromme}, \citenamefont {Barty},\ and\
  \citenamefont {Chapman}}]{pmid26863980}%
  \BibitemOpen
  \bibfield  {author} {\bibinfo {author} {\bibfnamefont {K.}~\bibnamefont
  {Ayyer}}, \bibinfo {author} {\bibfnamefont {O.~M.}\ \bibnamefont {Yefanov}},
  \bibinfo {author} {\bibfnamefont {D.}~\bibnamefont {Oberthur}}, \bibinfo
  {author} {\bibfnamefont {S.}~\bibnamefont {Roy-Chowdhury}}, \bibinfo {author}
  {\bibfnamefont {L.}~\bibnamefont {Galli}}, \bibinfo {author} {\bibfnamefont
  {V.}~\bibnamefont {Mariani}}, \bibinfo {author} {\bibfnamefont
  {S.}~\bibnamefont {Basu}}, \bibinfo {author} {\bibfnamefont {J.}~\bibnamefont
  {Coe}}, \bibinfo {author} {\bibfnamefont {C.~E.}\ \bibnamefont {Conrad}},
  \bibinfo {author} {\bibfnamefont {R.}~\bibnamefont {Fromme}}, \bibinfo
  {author} {\bibfnamefont {A.}~\bibnamefont {Schaffer}}, \bibinfo {author}
  {\bibfnamefont {K.}~\bibnamefont {Dorner}}, \bibinfo {author} {\bibfnamefont
  {D.}~\bibnamefont {James}}, \bibinfo {author} {\bibfnamefont
  {C.}~\bibnamefont {Kupitz}}, \bibinfo {author} {\bibfnamefont
  {M.}~\bibnamefont {Metz}}, \bibinfo {author} {\bibfnamefont {G.}~\bibnamefont
  {Nelson}}, \bibinfo {author} {\bibfnamefont {P.~L.}\ \bibnamefont {Xavier}},
  \bibinfo {author} {\bibfnamefont {K.~R.}\ \bibnamefont {Beyerlein}}, \bibinfo
  {author} {\bibfnamefont {M.}~\bibnamefont {Schmidt}}, \bibinfo {author}
  {\bibfnamefont {I.}~\bibnamefont {Sarrou}}, \bibinfo {author} {\bibfnamefont
  {J.~C.}\ \bibnamefont {Spence}}, \bibinfo {author} {\bibfnamefont
  {U.}~\bibnamefont {Weierstall}}, \bibinfo {author} {\bibfnamefont {T.~A.}\
  \bibnamefont {White}}, \bibinfo {author} {\bibfnamefont {J.~H.}\ \bibnamefont
  {Yang}}, \bibinfo {author} {\bibfnamefont {Y.}~\bibnamefont {Zhao}}, \bibinfo
  {author} {\bibfnamefont {M.}~\bibnamefont {Liang}}, \bibinfo {author}
  {\bibfnamefont {A.}~\bibnamefont {Aquila}}, \bibinfo {author} {\bibfnamefont
  {M.~S.}\ \bibnamefont {Hunter}}, \bibinfo {author} {\bibfnamefont {J.~S.}\
  \bibnamefont {Robinson}}, \bibinfo {author} {\bibfnamefont {J.~E.}\
  \bibnamefont {Koglin}}, \bibinfo {author} {\bibfnamefont {S.}~\bibnamefont
  {Boutet}}, \bibinfo {author} {\bibfnamefont {P.}~\bibnamefont {Fromme}},
  \bibinfo {author} {\bibfnamefont {A.}~\bibnamefont {Barty}}, \ and\ \bibinfo
  {author} {\bibfnamefont {H.~N.}\ \bibnamefont {Chapman}},\ }\href@noop {}
  {\bibfield  {journal} {\bibinfo  {journal} {Nature}\ }\textbf {\bibinfo
  {volume} {530}},\ \bibinfo {pages} {202} (\bibinfo {year}
  {2016})}\BibitemShut {NoStop}%
\bibitem [{\citenamefont {Meisburger}\ \emph {et~al.}(2017)\citenamefont
  {Meisburger}, \citenamefont {Thomas}, \citenamefont {Watkins},\ and\
  \citenamefont {Ando}}]{pmid28558231}%
  \BibitemOpen
  \bibfield  {author} {\bibinfo {author} {\bibfnamefont {S.~P.}\ \bibnamefont
  {Meisburger}}, \bibinfo {author} {\bibfnamefont {W.~C.}\ \bibnamefont
  {Thomas}}, \bibinfo {author} {\bibfnamefont {M.~B.}\ \bibnamefont {Watkins}},
  \ and\ \bibinfo {author} {\bibfnamefont {N.}~\bibnamefont {Ando}},\
  }\href@noop {} {\bibfield  {journal} {\bibinfo  {journal} {Chem. Rev.}\ ,\
  \bibinfo {pages} {7615}} (\bibinfo {year} {2017})}\BibitemShut {NoStop}%
\bibitem [{\citenamefont {Peck}\ \emph {et~al.}(2016)\citenamefont {Peck},
  \citenamefont {Sunden}, \citenamefont {Andrews}, \citenamefont {Pande},\ and\
  \citenamefont {Herschlag}}]{pmid27189921}%
  \BibitemOpen
  \bibfield  {author} {\bibinfo {author} {\bibfnamefont {A.}~\bibnamefont
  {Peck}}, \bibinfo {author} {\bibfnamefont {F.}~\bibnamefont {Sunden}},
  \bibinfo {author} {\bibfnamefont {L.~D.}\ \bibnamefont {Andrews}}, \bibinfo
  {author} {\bibfnamefont {V.~S.}\ \bibnamefont {Pande}}, \ and\ \bibinfo
  {author} {\bibfnamefont {D.}~\bibnamefont {Herschlag}},\ }\href@noop {}
  {\bibfield  {journal} {\bibinfo  {journal} {J. Mol. Biol.}\ }\textbf
  {\bibinfo {volume} {428}},\ \bibinfo {pages} {2758} (\bibinfo {year}
  {2016})}\BibitemShut {NoStop}%
\bibitem [{\citenamefont {Herrou}\ \emph {et~al.}(2016)\citenamefont {Herrou},
  \citenamefont {Czyz}, \citenamefont {Willett}, \citenamefont {Kim},
  \citenamefont {Chhor}, \citenamefont {Babnigg}, \citenamefont {Kim},\ and\
  \citenamefont {Crosson}}]{pmid26858101}%
  \BibitemOpen
  \bibfield  {author} {\bibinfo {author} {\bibfnamefont {J.}~\bibnamefont
  {Herrou}}, \bibinfo {author} {\bibfnamefont {D.~M.}\ \bibnamefont {Czyz}},
  \bibinfo {author} {\bibfnamefont {J.~W.}\ \bibnamefont {Willett}}, \bibinfo
  {author} {\bibfnamefont {H.~S.}\ \bibnamefont {Kim}}, \bibinfo {author}
  {\bibfnamefont {G.}~\bibnamefont {Chhor}}, \bibinfo {author} {\bibfnamefont
  {G.}~\bibnamefont {Babnigg}}, \bibinfo {author} {\bibfnamefont
  {Y.}~\bibnamefont {Kim}}, \ and\ \bibinfo {author} {\bibfnamefont
  {S.}~\bibnamefont {Crosson}},\ }\href@noop {} {\bibfield  {journal} {\bibinfo
   {journal} {J. Bacteriol.}\ }\textbf {\bibinfo {volume} {198}},\ \bibinfo
  {pages} {1281} (\bibinfo {year} {2016})}\BibitemShut {NoStop}%
\bibitem [{\citenamefont {Kabsch}(2010{\natexlab{a}})}]{pmid20124692}%
  \BibitemOpen
  \bibfield  {author} {\bibinfo {author} {\bibfnamefont {W.}~\bibnamefont
  {Kabsch}},\ }\href@noop {} {\bibfield  {journal} {\bibinfo  {journal} {Acta
  Crystallogr. D Biol. Crystallogr.}\ }\textbf {\bibinfo {volume} {66}},\
  \bibinfo {pages} {125} (\bibinfo {year} {2010}{\natexlab{a}})}\BibitemShut
  {NoStop}%
\bibitem [{\citenamefont {Hura}\ \emph {et~al.}(2000)\citenamefont {Hura},
  \citenamefont {Sorenson}, \citenamefont {Glaeser},\ and\ \citenamefont
  {Head-Gordon}}]{doi:10.1063/1.1319614}%
  \BibitemOpen
  \bibfield  {author} {\bibinfo {author} {\bibfnamefont {G.}~\bibnamefont
  {Hura}}, \bibinfo {author} {\bibfnamefont {J.~M.}\ \bibnamefont {Sorenson}},
  \bibinfo {author} {\bibfnamefont {R.~M.}\ \bibnamefont {Glaeser}}, \ and\
  \bibinfo {author} {\bibfnamefont {T.}~\bibnamefont {Head-Gordon}},\ }\href
  {\doibase 10.1063/1.1319614} {\bibfield  {journal} {\bibinfo  {journal} {The
  Journal of Chemical Physics}\ }\textbf {\bibinfo {volume} {113}},\ \bibinfo
  {pages} {9140} (\bibinfo {year} {2000})}\BibitemShut {NoStop}%
\bibitem [{\citenamefont {Wall}(2009)}]{pmid19488705}%
  \BibitemOpen
  \bibfield  {author} {\bibinfo {author} {\bibfnamefont {M.~E.}\ \bibnamefont
  {Wall}},\ }\href@noop {} {\bibfield  {journal} {\bibinfo  {journal} {Methods
  Mol. Biol.}\ }\textbf {\bibinfo {volume} {544}},\ \bibinfo {pages} {269}
  (\bibinfo {year} {2009})}\BibitemShut {NoStop}%
\bibitem [{\citenamefont {Waterman}\ \emph {et~al.}(2016)\citenamefont
  {Waterman}, \citenamefont {Winter}, \citenamefont {Gildea}, \citenamefont
  {Parkhurst}, \citenamefont {Brewster}, \citenamefont {Sauter},\ and\
  \citenamefont {Evans}}]{pmid27050135}%
  \BibitemOpen
  \bibfield  {author} {\bibinfo {author} {\bibfnamefont {D.~G.}\ \bibnamefont
  {Waterman}}, \bibinfo {author} {\bibfnamefont {G.}~\bibnamefont {Winter}},
  \bibinfo {author} {\bibfnamefont {R.~J.}\ \bibnamefont {Gildea}}, \bibinfo
  {author} {\bibfnamefont {J.~M.}\ \bibnamefont {Parkhurst}}, \bibinfo {author}
  {\bibfnamefont {A.~S.}\ \bibnamefont {Brewster}}, \bibinfo {author}
  {\bibfnamefont {N.~K.}\ \bibnamefont {Sauter}}, \ and\ \bibinfo {author}
  {\bibfnamefont {G.}~\bibnamefont {Evans}},\ }\href@noop {} {\bibfield
  {journal} {\bibinfo  {journal} {Acta Crystallogr D Struct Biol}\ }\textbf
  {\bibinfo {volume} {72}},\ \bibinfo {pages} {558} (\bibinfo {year}
  {2016})}\BibitemShut {NoStop}%
\bibitem [{\citenamefont {Boysen}\ and\ \citenamefont
  {Adlhart}(1987)}]{Boysen}%
  \BibitemOpen
  \bibfield  {author} {\bibinfo {author} {\bibfnamefont {H.}~\bibnamefont
  {Boysen}}\ and\ \bibinfo {author} {\bibfnamefont {W.}~\bibnamefont
  {Adlhart}},\ }\href {\doibase 10.1107/S0021889887086825} {\bibfield
  {journal} {\bibinfo  {journal} {Journal of Applied Crystallography}\ }\textbf
  {\bibinfo {volume} {20}},\ \bibinfo {pages} {200} (\bibinfo {year}
  {1987})}\BibitemShut {NoStop}%
\bibitem [{\citenamefont {Kabsch}(2010{\natexlab{b}})}]{pmid20124693}%
  \BibitemOpen
  \bibfield  {author} {\bibinfo {author} {\bibfnamefont {W.}~\bibnamefont
  {Kabsch}},\ }\href@noop {} {\bibfield  {journal} {\bibinfo  {journal} {Acta
  Crystallogr. D Biol. Crystallogr.}\ }\textbf {\bibinfo {volume} {66}},\
  \bibinfo {pages} {133} (\bibinfo {year} {2010}{\natexlab{b}})}\BibitemShut
  {NoStop}%
\bibitem [{\citenamefont {Polikanov}\ and\ \citenamefont
  {Moore}(2015)}]{pmid26457426}%
  \BibitemOpen
  \bibfield  {author} {\bibinfo {author} {\bibfnamefont {Y.~S.}\ \bibnamefont
  {Polikanov}}\ and\ \bibinfo {author} {\bibfnamefont {P.~B.}\ \bibnamefont
  {Moore}},\ }\href@noop {} {\bibfield  {journal} {\bibinfo  {journal} {Acta
  Crystallogr. D Biol. Crystallogr.}\ }\textbf {\bibinfo {volume} {71}},\
  \bibinfo {pages} {2021} (\bibinfo {year} {2015})}\BibitemShut {NoStop}%
\bibitem [{\citenamefont {Holton}(2016)}]{nonbragg}%
  \BibitemOpen
  \bibfield  {author} {\bibinfo {author} {\bibfnamefont {J.}~\bibnamefont
  {Holton}},\ }\href {http://bl831.als.lbl.gov/~jamesh/nonBragg} {\enquote
  {\bibinfo {title} {nonbragg},}\ } (\bibinfo {year} {2016})\BibitemShut
  {NoStop}%
\bibitem [{\citenamefont {McCoy}\ \emph {et~al.}(2007)\citenamefont {McCoy},
  \citenamefont {Grosse-Kunstleve}, \citenamefont {Adams}, \citenamefont
  {Winn}, \citenamefont {Storoni},\ and\ \citenamefont {Read}}]{pmid19461840}%
  \BibitemOpen
  \bibfield  {author} {\bibinfo {author} {\bibfnamefont {A.~J.}\ \bibnamefont
  {McCoy}}, \bibinfo {author} {\bibfnamefont {R.~W.}\ \bibnamefont
  {Grosse-Kunstleve}}, \bibinfo {author} {\bibfnamefont {P.~D.}\ \bibnamefont
  {Adams}}, \bibinfo {author} {\bibfnamefont {M.~D.}\ \bibnamefont {Winn}},
  \bibinfo {author} {\bibfnamefont {L.~C.}\ \bibnamefont {Storoni}}, \ and\
  \bibinfo {author} {\bibfnamefont {R.~J.}\ \bibnamefont {Read}},\ }\href@noop
  {} {\bibfield  {journal} {\bibinfo  {journal} {J Appl Crystallogr}\ }\textbf
  {\bibinfo {volume} {40}},\ \bibinfo {pages} {658} (\bibinfo {year}
  {2007})}\BibitemShut {NoStop}%
\bibitem [{\citenamefont {Adams}\ \emph {et~al.}(2010)\citenamefont {Adams},
  \citenamefont {Afonine}, \citenamefont {Bunkoczi}, \citenamefont {Chen},
  \citenamefont {Davis}, \citenamefont {Echols}, \citenamefont {Headd},
  \citenamefont {Hung}, \citenamefont {Kapral}, \citenamefont
  {Grosse-Kunstleve}, \citenamefont {McCoy}, \citenamefont {Moriarty},
  \citenamefont {Oeffner}, \citenamefont {Read}, \citenamefont {Richardson},
  \citenamefont {Richardson}, \citenamefont {Terwilliger},\ and\ \citenamefont
  {Zwart}}]{pmid20124702}%
  \BibitemOpen
  \bibfield  {author} {\bibinfo {author} {\bibfnamefont {P.~D.}\ \bibnamefont
  {Adams}}, \bibinfo {author} {\bibfnamefont {P.~V.}\ \bibnamefont {Afonine}},
  \bibinfo {author} {\bibfnamefont {G.}~\bibnamefont {Bunkoczi}}, \bibinfo
  {author} {\bibfnamefont {V.~B.}\ \bibnamefont {Chen}}, \bibinfo {author}
  {\bibfnamefont {I.~W.}\ \bibnamefont {Davis}}, \bibinfo {author}
  {\bibfnamefont {N.}~\bibnamefont {Echols}}, \bibinfo {author} {\bibfnamefont
  {J.~J.}\ \bibnamefont {Headd}}, \bibinfo {author} {\bibfnamefont {L.~W.}\
  \bibnamefont {Hung}}, \bibinfo {author} {\bibfnamefont {G.~J.}\ \bibnamefont
  {Kapral}}, \bibinfo {author} {\bibfnamefont {R.~W.}\ \bibnamefont
  {Grosse-Kunstleve}}, \bibinfo {author} {\bibfnamefont {A.~J.}\ \bibnamefont
  {McCoy}}, \bibinfo {author} {\bibfnamefont {N.~W.}\ \bibnamefont {Moriarty}},
  \bibinfo {author} {\bibfnamefont {R.}~\bibnamefont {Oeffner}}, \bibinfo
  {author} {\bibfnamefont {R.~J.}\ \bibnamefont {Read}}, \bibinfo {author}
  {\bibfnamefont {D.~C.}\ \bibnamefont {Richardson}}, \bibinfo {author}
  {\bibfnamefont {J.~S.}\ \bibnamefont {Richardson}}, \bibinfo {author}
  {\bibfnamefont {T.~C.}\ \bibnamefont {Terwilliger}}, \ and\ \bibinfo {author}
  {\bibfnamefont {P.~H.}\ \bibnamefont {Zwart}},\ }\href@noop {} {\bibfield
  {journal} {\bibinfo  {journal} {Acta Crystallogr. D Biol. Crystallogr.}\
  }\textbf {\bibinfo {volume} {66}},\ \bibinfo {pages} {213} (\bibinfo {year}
  {2010})}\BibitemShut {NoStop}%
\bibitem [{\citenamefont {Murshudov}\ \emph {et~al.}(2011)\citenamefont
  {Murshudov}, \citenamefont {Skubak}, \citenamefont {Lebedev}, \citenamefont
  {Pannu}, \citenamefont {Steiner}, \citenamefont {Nicholls}, \citenamefont
  {Winn}, \citenamefont {Long},\ and\ \citenamefont {Vagin}}]{pmid21460454}%
  \BibitemOpen
  \bibfield  {author} {\bibinfo {author} {\bibfnamefont {G.~N.}\ \bibnamefont
  {Murshudov}}, \bibinfo {author} {\bibfnamefont {P.}~\bibnamefont {Skubak}},
  \bibinfo {author} {\bibfnamefont {A.~A.}\ \bibnamefont {Lebedev}}, \bibinfo
  {author} {\bibfnamefont {N.~S.}\ \bibnamefont {Pannu}}, \bibinfo {author}
  {\bibfnamefont {R.~A.}\ \bibnamefont {Steiner}}, \bibinfo {author}
  {\bibfnamefont {R.~A.}\ \bibnamefont {Nicholls}}, \bibinfo {author}
  {\bibfnamefont {M.~D.}\ \bibnamefont {Winn}}, \bibinfo {author}
  {\bibfnamefont {F.}~\bibnamefont {Long}}, \ and\ \bibinfo {author}
  {\bibfnamefont {A.~A.}\ \bibnamefont {Vagin}},\ }\href@noop {} {\bibfield
  {journal} {\bibinfo  {journal} {Acta Crystallogr. D Biol. Crystallogr.}\
  }\textbf {\bibinfo {volume} {67}},\ \bibinfo {pages} {355} (\bibinfo {year}
  {2011})}\BibitemShut {NoStop}%
\bibitem [{\citenamefont {Emsley}\ \emph {et~al.}(2010)\citenamefont {Emsley},
  \citenamefont {Lohkamp}, \citenamefont {Scott},\ and\ \citenamefont
  {Cowtan}}]{pmid20383002}%
  \BibitemOpen
  \bibfield  {author} {\bibinfo {author} {\bibfnamefont {P.}~\bibnamefont
  {Emsley}}, \bibinfo {author} {\bibfnamefont {B.}~\bibnamefont {Lohkamp}},
  \bibinfo {author} {\bibfnamefont {W.~G.}\ \bibnamefont {Scott}}, \ and\
  \bibinfo {author} {\bibfnamefont {K.}~\bibnamefont {Cowtan}},\ }\href@noop {}
  {\bibfield  {journal} {\bibinfo  {journal} {Acta Crystallogr. D Biol.
  Crystallogr.}\ }\textbf {\bibinfo {volume} {66}},\ \bibinfo {pages} {486}
  (\bibinfo {year} {2010})}\BibitemShut {NoStop}%
\bibitem [{\citenamefont {Fraser}(2015)}]{sbdb68}%
  \BibitemOpen
  \bibfield  {author} {\bibinfo {author} {\bibfnamefont {J.}~\bibnamefont
  {Fraser}},\ }\href {\doibase 10.15785/SBGRID/68} {\enquote {\bibinfo {title}
  {X-ray diffraction data for: Cyclophilin a. pdb code 4yuo},}\ } (\bibinfo
  {year} {2015})\BibitemShut {NoStop}%
\bibitem [{\citenamefont {Peck}\ \emph {et~al.}(2017)\citenamefont {Peck},
  \citenamefont {Ressl},\ and\ \citenamefont {Herschlag}}]{sbdb456}%
  \BibitemOpen
  \bibfield  {author} {\bibinfo {author} {\bibfnamefont {A.}~\bibnamefont
  {Peck}}, \bibinfo {author} {\bibfnamefont {S.}~\bibnamefont {Ressl}}, \ and\
  \bibinfo {author} {\bibfnamefont {D.}~\bibnamefont {Herschlag}},\ }\href
  {\doibase 10.15785/SBGRID/456} {\enquote {\bibinfo {title} {X-ray diffraction
  data for: E. coli alkaline phosphatase in complex with tungstate. pdb code
  5c66},}\ } (\bibinfo {year} {2017})\BibitemShut {NoStop}%
\bibitem [{\citenamefont {Herrou}\ and\ \citenamefont
  {Crosson}(2015)}]{sbdb203}%
  \BibitemOpen
  \bibfield  {author} {\bibinfo {author} {\bibfnamefont {J.}~\bibnamefont
  {Herrou}}\ and\ \bibinfo {author} {\bibfnamefont {S.}~\bibnamefont
  {Crosson}},\ }\href {\doibase 10.15785/SBGRID/203} {\enquote {\bibinfo
  {title} {X-ray diffraction data for: Structure of b. abortus wrba-related
  protein a (apo-wrpa). pdb code 5f51},}\ } (\bibinfo {year}
  {2015})\BibitemShut {NoStop}%
\bibitem [{\citenamefont {Bray}\ \emph {et~al.}(2011)\citenamefont {Bray},
  \citenamefont {Weiss},\ and\ \citenamefont {Levitt}}]{pmid22208195}%
  \BibitemOpen
  \bibfield  {author} {\bibinfo {author} {\bibfnamefont {J.~K.}\ \bibnamefont
  {Bray}}, \bibinfo {author} {\bibfnamefont {D.~R.}\ \bibnamefont {Weiss}}, \
  and\ \bibinfo {author} {\bibfnamefont {M.}~\bibnamefont {Levitt}},\
  }\href@noop {} {\bibfield  {journal} {\bibinfo  {journal} {Biophys. J.}\
  }\textbf {\bibinfo {volume} {101}},\ \bibinfo {pages} {2966} (\bibinfo {year}
  {2011})}\BibitemShut {NoStop}%
\bibitem [{\citenamefont {Guinier}(1963)}]{guinier}%
  \BibitemOpen
  \bibfield  {author} {\bibinfo {author} {\bibfnamefont {A.}~\bibnamefont
  {Guinier}},\ }\href@noop {} {\emph {\bibinfo {title} {X-ray diffraction in
  crystals, imperfect crystals, and amorphous bodies}}}\ (\bibinfo  {publisher}
  {W. H. Freeman, San Francisco},\ \bibinfo {year} {1963})\BibitemShut
  {NoStop}%
\bibitem [{\citenamefont {Clarage}\ \emph {et~al.}(1992)\citenamefont
  {Clarage}, \citenamefont {Clarage}, \citenamefont {Phillips}, \citenamefont
  {Sweet},\ and\ \citenamefont {Caspar}}]{pmid1603804}%
  \BibitemOpen
  \bibfield  {author} {\bibinfo {author} {\bibfnamefont {J.~B.}\ \bibnamefont
  {Clarage}}, \bibinfo {author} {\bibfnamefont {M.~S.}\ \bibnamefont
  {Clarage}}, \bibinfo {author} {\bibfnamefont {W.~C.}\ \bibnamefont
  {Phillips}}, \bibinfo {author} {\bibfnamefont {R.~M.}\ \bibnamefont {Sweet}},
  \ and\ \bibinfo {author} {\bibfnamefont {D.~L.}\ \bibnamefont {Caspar}},\
  }\href@noop {} {\bibfield  {journal} {\bibinfo  {journal} {Proteins}\
  }\textbf {\bibinfo {volume} {12}},\ \bibinfo {pages} {145} (\bibinfo {year}
  {1992})}\BibitemShut {NoStop}%
\bibitem [{\citenamefont {Lane}(2017)}]{thor}%
  \BibitemOpen
  \bibfield  {author} {\bibinfo {author} {\bibfnamefont {T.~J.}\ \bibnamefont
  {Lane}},\ }\href@noop {} {\enquote {\bibinfo {title} {Thor},}\ }\bibinfo
  {howpublished} {\url{https://github.com/tjlane/thor}} (\bibinfo {year}
  {2017})\BibitemShut {NoStop}%
\bibitem [{\citenamefont {Karplus}\ and\ \citenamefont
  {Diederichs}(2012)}]{pmid22628654}%
  \BibitemOpen
  \bibfield  {author} {\bibinfo {author} {\bibfnamefont {P.~A.}\ \bibnamefont
  {Karplus}}\ and\ \bibinfo {author} {\bibfnamefont {K.}~\bibnamefont
  {Diederichs}},\ }\href@noop {} {\bibfield  {journal} {\bibinfo  {journal}
  {Science}\ }\textbf {\bibinfo {volume} {336}},\ \bibinfo {pages} {1030}
  (\bibinfo {year} {2012})}\BibitemShut {NoStop}%
\bibitem [{\citenamefont {van~den Bedem}\ \emph {et~al.}(2009)\citenamefont
  {van~den Bedem}, \citenamefont {Dhanik}, \citenamefont {Latombe},\ and\
  \citenamefont {Deacon}}]{pmid19770508}%
  \BibitemOpen
  \bibfield  {author} {\bibinfo {author} {\bibfnamefont {H.}~\bibnamefont
  {van~den Bedem}}, \bibinfo {author} {\bibfnamefont {A.}~\bibnamefont
  {Dhanik}}, \bibinfo {author} {\bibfnamefont {J.~C.}\ \bibnamefont {Latombe}},
  \ and\ \bibinfo {author} {\bibfnamefont {A.~M.}\ \bibnamefont {Deacon}},\
  }\href@noop {} {\bibfield  {journal} {\bibinfo  {journal} {Acta Crystallogr.
  D Biol. Crystallogr.}\ }\textbf {\bibinfo {volume} {65}},\ \bibinfo {pages}
  {1107} (\bibinfo {year} {2009})}\BibitemShut {NoStop}%
\bibitem [{\citenamefont {Fraser}\ \emph {et~al.}(2009)\citenamefont {Fraser},
  \citenamefont {Clarkson}, \citenamefont {Degnan}, \citenamefont {Erion},
  \citenamefont {Kern},\ and\ \citenamefont {Alber}}]{pmid19956261}%
  \BibitemOpen
  \bibfield  {author} {\bibinfo {author} {\bibfnamefont {J.~S.}\ \bibnamefont
  {Fraser}}, \bibinfo {author} {\bibfnamefont {M.~W.}\ \bibnamefont
  {Clarkson}}, \bibinfo {author} {\bibfnamefont {S.~C.}\ \bibnamefont
  {Degnan}}, \bibinfo {author} {\bibfnamefont {R.}~\bibnamefont {Erion}},
  \bibinfo {author} {\bibfnamefont {D.}~\bibnamefont {Kern}}, \ and\ \bibinfo
  {author} {\bibfnamefont {T.}~\bibnamefont {Alber}},\ }\href@noop {}
  {\bibfield  {journal} {\bibinfo  {journal} {Nature}\ }\textbf {\bibinfo
  {volume} {462}},\ \bibinfo {pages} {669} (\bibinfo {year}
  {2009})}\BibitemShut {NoStop}%
\bibitem [{\citenamefont {Fraser}\ \emph {et~al.}(2011)\citenamefont {Fraser},
  \citenamefont {van~den Bedem}, \citenamefont {Samelson}, \citenamefont
  {Lang}, \citenamefont {Holton}, \citenamefont {Echols},\ and\ \citenamefont
  {Alber}}]{pmid21918110}%
  \BibitemOpen
  \bibfield  {author} {\bibinfo {author} {\bibfnamefont {J.~S.}\ \bibnamefont
  {Fraser}}, \bibinfo {author} {\bibfnamefont {H.}~\bibnamefont {van~den
  Bedem}}, \bibinfo {author} {\bibfnamefont {A.~J.}\ \bibnamefont {Samelson}},
  \bibinfo {author} {\bibfnamefont {P.~T.}\ \bibnamefont {Lang}}, \bibinfo
  {author} {\bibfnamefont {J.~M.}\ \bibnamefont {Holton}}, \bibinfo {author}
  {\bibfnamefont {N.}~\bibnamefont {Echols}}, \ and\ \bibinfo {author}
  {\bibfnamefont {T.}~\bibnamefont {Alber}},\ }\href@noop {} {\bibfield
  {journal} {\bibinfo  {journal} {Proc. Natl. Acad. Sci. U.S.A.}\ }\textbf
  {\bibinfo {volume} {108}},\ \bibinfo {pages} {16247} (\bibinfo {year}
  {2011})}\BibitemShut {NoStop}%
\bibitem [{\citenamefont {Keedy}\ \emph {et~al.}(2015)\citenamefont {Keedy},
  \citenamefont {Kenner}, \citenamefont {Warkentin}, \citenamefont {Woldeyes},
  \citenamefont {Hopkins}, \citenamefont {Thompson}, \citenamefont {Brewster},
  \citenamefont {Van~Benschoten}, \citenamefont {Baxter}, \citenamefont
  {Uervirojnangkoorn}, \citenamefont {McPhillips}, \citenamefont {Song},
  \citenamefont {Alonso-Mori}, \citenamefont {Holton}, \citenamefont {Weis},
  \citenamefont {Brunger}, \citenamefont {Soltis}, \citenamefont {Lemke},
  \citenamefont {Gonzalez}, \citenamefont {Sauter}, \citenamefont {Cohen},
  \citenamefont {van~den Bedem}, \citenamefont {Thorne},\ and\ \citenamefont
  {Fraser}}]{pmid26422513}%
  \BibitemOpen
  \bibfield  {author} {\bibinfo {author} {\bibfnamefont {D.~A.}\ \bibnamefont
  {Keedy}}, \bibinfo {author} {\bibfnamefont {L.~R.}\ \bibnamefont {Kenner}},
  \bibinfo {author} {\bibfnamefont {M.}~\bibnamefont {Warkentin}}, \bibinfo
  {author} {\bibfnamefont {R.~A.}\ \bibnamefont {Woldeyes}}, \bibinfo {author}
  {\bibfnamefont {J.~B.}\ \bibnamefont {Hopkins}}, \bibinfo {author}
  {\bibfnamefont {M.~C.}\ \bibnamefont {Thompson}}, \bibinfo {author}
  {\bibfnamefont {A.~S.}\ \bibnamefont {Brewster}}, \bibinfo {author}
  {\bibfnamefont {A.~H.}\ \bibnamefont {Van~Benschoten}}, \bibinfo {author}
  {\bibfnamefont {E.~L.}\ \bibnamefont {Baxter}}, \bibinfo {author}
  {\bibfnamefont {M.}~\bibnamefont {Uervirojnangkoorn}}, \bibinfo {author}
  {\bibfnamefont {S.~E.}\ \bibnamefont {McPhillips}}, \bibinfo {author}
  {\bibfnamefont {J.}~\bibnamefont {Song}}, \bibinfo {author} {\bibfnamefont
  {R.}~\bibnamefont {Alonso-Mori}}, \bibinfo {author} {\bibfnamefont {J.~M.}\
  \bibnamefont {Holton}}, \bibinfo {author} {\bibfnamefont {W.~I.}\
  \bibnamefont {Weis}}, \bibinfo {author} {\bibfnamefont {A.~T.}\ \bibnamefont
  {Brunger}}, \bibinfo {author} {\bibfnamefont {S.~M.}\ \bibnamefont {Soltis}},
  \bibinfo {author} {\bibfnamefont {H.}~\bibnamefont {Lemke}}, \bibinfo
  {author} {\bibfnamefont {A.}~\bibnamefont {Gonzalez}}, \bibinfo {author}
  {\bibfnamefont {N.~K.}\ \bibnamefont {Sauter}}, \bibinfo {author}
  {\bibfnamefont {A.~E.}\ \bibnamefont {Cohen}}, \bibinfo {author}
  {\bibfnamefont {H.}~\bibnamefont {van~den Bedem}}, \bibinfo {author}
  {\bibfnamefont {R.~E.}\ \bibnamefont {Thorne}}, \ and\ \bibinfo {author}
  {\bibfnamefont {J.~S.}\ \bibnamefont {Fraser}},\ }\href@noop {} {\bibfield
  {journal} {\bibinfo  {journal} {Elife}\ }\textbf {\bibinfo {volume} {4}}
  (\bibinfo {year} {2015})}\BibitemShut {NoStop}%
\bibitem [{\citenamefont {Wilson}(2013)}]{pmid23985728}%
  \BibitemOpen
  \bibfield  {author} {\bibinfo {author} {\bibfnamefont {M.~A.}\ \bibnamefont
  {Wilson}},\ }\href@noop {} {\bibfield  {journal} {\bibinfo  {journal} {Nat.
  Methods}\ }\textbf {\bibinfo {volume} {10}},\ \bibinfo {pages} {835}
  (\bibinfo {year} {2013})}\BibitemShut {NoStop}%
\bibitem [{\citenamefont {Glover}\ \emph {et~al.}(1991)\citenamefont {Glover},
  \citenamefont {Harris}, \citenamefont {Helliwell},\ and\ \citenamefont
  {Moss}}]{bw0632}%
  \BibitemOpen
  \bibfield  {author} {\bibinfo {author} {\bibfnamefont {I.~D.}\ \bibnamefont
  {Glover}}, \bibinfo {author} {\bibfnamefont {G.~W.}\ \bibnamefont {Harris}},
  \bibinfo {author} {\bibfnamefont {J.~R.}\ \bibnamefont {Helliwell}}, \ and\
  \bibinfo {author} {\bibfnamefont {D.~S.}\ \bibnamefont {Moss}},\ }\href
  {\doibase 10.1107/S0108768191004585} {\bibfield  {journal} {\bibinfo
  {journal} {Acta Crystallographica Section B}\ }\textbf {\bibinfo {volume}
  {47}},\ \bibinfo {pages} {960} (\bibinfo {year} {1991})}\BibitemShut
  {NoStop}%
\bibitem [{\citenamefont {Meinhold}\ \emph {et~al.}(2007)\citenamefont
  {Meinhold}, \citenamefont {Merzel},\ and\ \citenamefont
  {Smith}}]{pmid17930640}%
  \BibitemOpen
  \bibfield  {author} {\bibinfo {author} {\bibfnamefont {L.}~\bibnamefont
  {Meinhold}}, \bibinfo {author} {\bibfnamefont {F.}~\bibnamefont {Merzel}}, \
  and\ \bibinfo {author} {\bibfnamefont {J.~C.}\ \bibnamefont {Smith}},\
  }\href@noop {} {\bibfield  {journal} {\bibinfo  {journal} {Phys. Rev. Lett.}\
  }\textbf {\bibinfo {volume} {99}},\ \bibinfo {pages} {138101} (\bibinfo
  {year} {2007})}\BibitemShut {NoStop}%
\bibitem [{\citenamefont {Doucet}\ and\ \citenamefont
  {Benoit}(1987)}]{pmid3808065}%
  \BibitemOpen
  \bibfield  {author} {\bibinfo {author} {\bibfnamefont {J.}~\bibnamefont
  {Doucet}}\ and\ \bibinfo {author} {\bibfnamefont {J.~P.}\ \bibnamefont
  {Benoit}},\ }\href@noop {} {\bibfield  {journal} {\bibinfo  {journal}
  {Nature}\ }\textbf {\bibinfo {volume} {325}},\ \bibinfo {pages} {643}
  (\bibinfo {year} {1987})}\BibitemShut {NoStop}%
\bibitem [{\citenamefont {Sayre}(1952)}]{Sayre:a00763}%
  \BibitemOpen
  \bibfield  {author} {\bibinfo {author} {\bibfnamefont {D.}~\bibnamefont
  {Sayre}},\ }\href {\doibase 10.1107/S0365110X52002276} {\bibfield  {journal}
  {\bibinfo  {journal} {Acta Crystallographica}\ }\textbf {\bibinfo {volume}
  {5}},\ \bibinfo {pages} {843} (\bibinfo {year} {1952})}\BibitemShut {NoStop}%
\bibitem [{\citenamefont {Faure}\ \emph {et~al.}(1994)\citenamefont {Faure},
  \citenamefont {Micu}, \citenamefont {Perahia}, \citenamefont {Doucet},
  \citenamefont {Smith},\ and\ \citenamefont {Benoit}}]{pmid7656016}%
  \BibitemOpen
  \bibfield  {author} {\bibinfo {author} {\bibfnamefont {P.}~\bibnamefont
  {Faure}}, \bibinfo {author} {\bibfnamefont {A.}~\bibnamefont {Micu}},
  \bibinfo {author} {\bibfnamefont {D.}~\bibnamefont {Perahia}}, \bibinfo
  {author} {\bibfnamefont {J.}~\bibnamefont {Doucet}}, \bibinfo {author}
  {\bibfnamefont {J.~C.}\ \bibnamefont {Smith}}, \ and\ \bibinfo {author}
  {\bibfnamefont {J.~P.}\ \bibnamefont {Benoit}},\ }\href@noop {} {\bibfield
  {journal} {\bibinfo  {journal} {Nat. Struct. Biol.}\ }\textbf {\bibinfo
  {volume} {1}},\ \bibinfo {pages} {124} (\bibinfo {year} {1994})}\BibitemShut
  {NoStop}%
\end{thebibliography}%


\begin{thebibliography}{11}
\bibitem{s_guinier} A. Guinier, \textit{X-ray diffraction in crystals, imperfect crystals, and amorphous bodies} (W. H. Freeman, San Francisco, 1963).
\bibitem{s_moore2009} P. B. Moore, Structure \textbf{17}, 1307 (2009)
\bibitem{s_chapman} K. Ayyer, O. M. Yefanov, D. Oberthur, S. RoyChowdhury, L. Galli, V. Mariani, S. Basu, J. Coe, C. E. Conrad, R. Fromme, A. Schaffer, K. Dorner, D. James, C. Kupitz, M. Metz, G. Nelson, P. L. Xavier, K. R. Beyerlein, M. Schmidt, I. Sarrou, J. C. Spence, U. Weierstall, T. A. White, J. H. Yang, Y. Zhao, M. Liang, A. Aquila, M. S. Hunter, J. S. Robinson, J. E. Koglin, S. Boutet, P. Fromme, A. Barty, and H. N. Chapman, Nature \textbf{530}, 202 (2016).
\bibitem{s_wall1997} M. E. Wall, J. B. Clarage, and G. N. Phillips, Structure \textbf{5}, 1599 (1997).
\bibitem{s_clarage} J. B. Clarage, M. S. Clarage, W. C. Phillips, R. M. Sweet, and D. L. Caspar, Proteins \textbf{12}, 145 (1992).
\bibitem{s_levitt} J. K. Bray, D. R. Weiss, and M. Levitt, Biophys. J. \textbf{101}, 2966 (2011).
\end{thebibliography}

\widetext
\clearpage
\begin{center}
\textbf{\large Supplemental Materials}
\end{center}
\setcounter{equation}{0}
\setcounter{figure}{0}
\setcounter{table}{0}
\setcounter{page}{1}
\setcounter{section}{0}
\makeatletter
\renewcommand{\theequation}{S\arabic{equation}}
\renewcommand{\thefigure}{S\arabic{figure}}
\renewcommand{\bibnumfmt}[1]{[S#1]}
\renewcommand{\citenumfont}[1]{S#1}

\section{Extended Methods}

The disorder models assessed in this study are described in more detail below. With the exception of the traditional model of liquid-like motions, in which correlations extend throughout the crystal, these models assume that correlations are confined within the boundaries of the asymmetric unit. Because of this assumption, the models of rigid body disorder and liquid-like motions without crystal neighbors share the symmetrized molecular transform, $I_{\mathrm{m}}$, as their basis:
\begin{equation}
I_{\mathrm{m}} = \sum\limits_{\textrm{asu}} | F_{\textrm{asu}}(\mathbf{q}) |^2
\label{eq:mtransform}
\end{equation}
where the summation is across all asymmetric units in the unit cell. Thus, this subset of models predicts that the diffuse scattering is a blurred image of the symmetrized molecular transform, with the form of the blurring dependent on the nature of the disorder. This incoherent sum of the asymmetric unit intensities is distinct from the coherent sum of scattered intensities across the crystal or, equivalently, the crystal transform, $I_{\mathrm{c}}$:
\begin{equation}
I_{\mathrm{c}} = | S (\mathbf{q}) |^2 \left| \sum\limits_{\textrm{asu}} F_{\textrm{asu}}(\mathbf{q}) ^2 \right|
\label{eq:uctransform}
\end{equation}
where the Dirac comb, $| S (\mathbf{q}) |^2$, is a nonzero constant at integral Miller indices and zero at all other $\mathbf{q}$. This is in contrast to the molecular transform, which is characterized by positive, nonzero intensities throughout reciprocal space. This distinction is important for the two variants of liquid-like motions models considered in this work, as noted below.

What follows is a derivation of a general expression for diffuse scattering in the Gaussian approximation, a category that encompasses the elastic network, rigid body translations, and liquid-like motions models described below. An expression for the total ensemble (or in ergodic systems, time) averaged scattered intensity at some wavevector $\mathbf{q}$ is:
\begin{equation}
\langle I (\mathbf{q}) \rangle = \left\langle \sum\limits_{c,d} e^{-i \mathbf{q} \cdot \mathbf{u}_{c d}} \sum\limits_{i,j} f_i f_j e^{-i \mathbf{q} \cdot \mathbf{r}_{ij}} e^{-i \mathbf{q} \cdot \boldsymbol{\delta}_{c_i d_j}} \right\rangle 
\label{eq:avgscatter}
\end{equation}
where $f_i$ is the atomic form factor for atom $i$, $\mathbf{u}$$_{cd}$ is the vector between the origins of unit cells $c$ and $d$, $\mathbf{r}$$_{ij}$ is the interatomic distance vector between atoms $i$ and $j$, and $\boldsymbol{\delta}$$_{c_i d_j}$ is the interatomic displacement vector between atom $i$ in unit cell $c$ and atom $j$ in unit cell $d$. In the Gaussian approximation, the statistical ensemble of atomic displacements may be described by a pairwise multivariate normal distribution, with zero mean and covariance matrix, $V_{c_id_j} = \langle \boldsymbol{\delta}_{c_i}^T\boldsymbol{\delta}_{d_j} \rangle \in \mathcal{R}^{3 \times 3}$:
\[
p( \boldsymbol{\delta}_{c_i}, \boldsymbol{\delta}_{d_j} ) \sim \mathrm{MVN}( \mathbf{0}, V_{c_id_j})
\]
Since the average is over pairwise probability distributions, eq.~\ref{eq:avgscatter} may be rewritten:
\begin{align}
\langle I (\mathbf{q}) \rangle &= \sum\limits_{c, d} e^{-i \mathbf{q} \cdot \mathbf{u}_{c d}} 
\sum\limits_{i,j} f_i f_j e^{-i \mathbf{q} \cdot \mathbf{r}_{ij}} \iint p( \boldsymbol{\delta}_{c_i},  \boldsymbol{\delta}_{d_j} )
e^{-i \mathbf{q} \cdot ( \boldsymbol{\delta}_{c_i}  - \boldsymbol{\delta}_{d_j})} 
d \boldsymbol{\delta}_{c_i}  \, d \boldsymbol{\delta}_{d_j}
\nonumber \\
&= \sum\limits_{c, d} e^{-i \mathbf{q} \cdot \mathbf{u}_{c d}} 
\sum\limits_{i,j} f_i f_j e^{-i \mathbf{q} \cdot \mathbf{r}_{ij}}
e^{- \frac{1}{2} \mathbf{q}^T V_{c_i c_i} \mathbf{q}
- \frac{1}{2} \mathbf{q}^T V_{d_j d_j} \mathbf{q}
+ \mathbf{q}^T V_{c_i d_j} \mathbf{q}}
\label{eq:mvnscatter}
\end{align}
We further assume that:
\begin{enumerate}
\item Correlations between atomic displacements in different unit cells are independent: $\langle \boldsymbol{\delta}_{c_i}^T \boldsymbol{\delta}_{d_j} \rangle = \mathbf{0} 
\ \mathrm{if} \ c \neq d$.
\item Atoms in different unit cells behave identically in a statistical fashion: $p(\boldsymbol{\delta}_{c i}) = p(\boldsymbol{\delta}_{d i})$ for all $i$.
\end{enumerate}
With these simplifying assumptions, $V_{c_i d_j} = \mathbf{0}$ if $c \neq d$, and $V_{c_i c_i}$ is identical for all $c$, such that $V_{c_i c_i} = V_{ii}$ (and similarly $V_{d_j d_j} = V_{jj}$). Eq.~\ref{eq:mvnscatter} can then be split into two terms: one expressing interference between unit cells (where $V_{c_i d_j} = \mathbf{0}$), and one expressing interference \textit{within} repeats of a single cell:
\begin{align}
\langle I (\mathbf{q}) \rangle =&
%
\sum\limits_{c,d \neq c} e^{-i \mathbf{q} \cdot \mathbf{u}_{c d}} 
\sum\limits_{i,j} f_i f_j e^{-i \mathbf{q} \cdot \mathbf{r}_{ij}} 
e^{
- \frac{1}{2} \mathbf{q}^T V_{ii} \mathbf{q}
- \frac{1}{2} \mathbf{q}^T V_{jj} \mathbf{q}
} \nonumber \\
%
&+
N \sum\limits_{i,j} f_i f_j e^{-i \mathbf{q} \cdot \mathbf{r}_{ij}} 
e^{
- \frac{1}{2} \mathbf{q}^T V_{ii} \mathbf{q}
- \frac{1}{2} \mathbf{q}^T V_{jj} \mathbf{q}
+ \mathbf{q}^T V_{ij} \mathbf{q}
}  \\
=&
%
\sum\limits_{c, d} e^{-i \mathbf{q} \cdot \mathbf{u}_{c d}} 
\sum\limits_{i,j} f_i f_j e^{-i \mathbf{q} \cdot \mathbf{r}_{ij}} 
e^{
- \frac{1}{2} \mathbf{q}^T V_{ii} \mathbf{q}
- \frac{1}{2} \mathbf{q}^T V_{jj} \mathbf{q}
} \nonumber \\
%
&+
N \sum\limits_{i,j} f_i f_j e^{-i \mathbf{q} \cdot \mathbf{r}_{ij}} 
e^{
- \frac{1}{2} \mathbf{q}^T V_{ii} \mathbf{q}
- \frac{1}{2} \mathbf{q}^T V_{jj} \mathbf{q}
}
\left[ 
e^{\mathbf{q}^T V_{ij} \mathbf{q}} - 1
\right]
\end{align}
where $N$ is the number of unit cells. The first term is recognizable as the expression corresponding to Bragg diffraction:
\begin{align}
I(\mathbf{q})_{\mathrm{Bragg}} &=
\sum\limits_{c, d} e^{-i \mathbf{q} \cdot \mathbf{u}_{c d}} 
\sum\limits_{i,j} f_i f_j e^{-i \mathbf{q} \cdot \mathbf{r}_{ij}} 
e^{
- \frac{1}{2} \mathbf{q}^T V_{ii} \mathbf{q}
- \frac{1}{2} \mathbf{q}^T V_{jj} \mathbf{q}
} \\
&= \left| \left( 
\sum_c e^{-i \mathbf{q} \cdot \mathbf{u}_{c}} 
\right) \left(
\sum\limits_{i} f_i e^{-i \mathbf{q} \cdot \mathbf{r}_{i}} 
e^{
- \frac{1}{2} \mathbf{q}^T V_{ii} \mathbf{q}
} \right) \right|^2\\
&= \left| S( \mathbf{q} ) \right|^2  \left| F( \mathbf{q} ) \right|^2
\end{align}
where atomic form factors are scaled by anisotropic Debye-Waller factors. $\left| S( \mathbf{q} ) \right|^2$ becomes a Dirac comb as the number of unit cells grows, showing this scattering is localized to discrete regions of $\mathbf{q}$. We consider the remaining scattering $\langle I (\mathbf{q}) \rangle - I(\mathbf{q})_{\mathrm{Bragg}}$ to be the diffuse scattering intensity:
\begin{equation}
I_\mathrm{diffuse} (\mathbf{q}) = N \sum\limits_{i,j} f_i f_j e^{-i \mathbf{q} \cdot \mathbf{r}_{ij}} 
e^{
- \frac{1}{2} \mathbf{q}^T V_{ii} \mathbf{q}
- \frac{1}{2} \mathbf{q}^T V_{jj} \mathbf{q}
}
\left[ 
e^{\mathbf{q}^T V_{ij} \mathbf{q}} - 1
\right]
\label{eq:diffuse}
\end{equation}
There are two notable features of the diffuse scattering. First, lacking the lattice transform $\left| S( \mathbf{q} ) \right|^2$ it is \textit{not} localized in reciprocal space. Second, it is non-trivial only if there are correlated displacements between atoms, i.e. when $V_{ij} \neq \mathbf{0}$.

\subsection*{Elastic network model} This model makes the assumptions outlined above in deriving eq. \ref{eq:diffuse}, with the further restriction that correlations be confined within the boundaries of the asymmetric unit. This additional assumption renders this model more biologically interpretable. For each system, the covariance matrix, $V_{ij}$, was determined from an elastic network model of the ordered atoms in the asymmetric unit. Specifically, the normal modes of each system were generated based on the protein's topology in torsion angle space, with a uniform spring constant for all atom pairs within a certain distance in this internal coordinate space \cite{s_levitt}. The first ten normal modes were then summed to generate $C_{ij}$, the isotropic correlation coefficient between the displacements of asymmetric unit atoms $i$ and $j$. Entries in this correlation matrix were converted to covariances using the following formula:
\begin{equation}
V_{ij} = C_{ij}\sqrt{ \langle \delta_i^2 \rangle \langle \delta_j^2 \rangle }
\label{eq:llm_cov}
\end{equation}
where the mean-square atomic displacements, $\langle \delta_i^2 \rangle$, are related to the isotropic B factors by: $B_i = 8 \pi^2 \langle \delta_i^2 \rangle$. Thus, the amplitudes of motions described by the covariance matrix are consistent with the refined Bragg models. Diffuse scattering maps were predicted from these covariance matrices using eq.~\ref{eq:diffuse}. 

\subsection*{Rigid body translational disorder}
The model of rigid body translational disorder is a special case of diffuse scattering in the Gaussian approximation (eq.~\ref{eq:diffuse}) that further assumes that all atoms in the asymmetric unit are displaced as a rigid structural unit. The displacement covariance between all atom pairs is thus identical and assuming isotropic translations can be described by a scalar, $\sigma^2$. The expression for the diffuse scattering intensity is:
\begin{align}
I_\mathrm{diffuse} (\mathbf{q}) =&\
N \sum\limits_{i,j} f_i f_j e^{-i \mathbf{q} \cdot \mathbf{r}_{ij}} 
\left[ 1 - e^{- q^2 \sigma^2} \right] 
\nonumber \\ 
=&\ N\left[ 1 - e^{- q^2 \sigma^2} \right] I_\mathrm{m}
\label{eq:chapman}
\end{align}
where $N$ is the number of unit cells and $I_m$ is the symmetrized molecular transform (eq.~\ref{eq:mtransform}). This expression has previously been derived in Refs. \cite{s_moore2009} and \cite{s_chapman}. For each system, the molecular transform was computed from the refined Bragg coordinates, excluding solvent, hydrogen, and crystallographically-unresolved atoms (which were assumed to exhibit uncorrelated disordered behavior and thus contribute to radially symmetric rather than anisotropic diffuse scattering). Best fit values of $\sigma$ were determined by scanning over this parameter to maximize the CC between the experimental and predicted maps. 

\subsection*{Liquid-like motions}
The liquid-like motions model is a specific case of Gaussian translational disorder in which correlated motions between atoms decay with interatomic distance \cite{s_wall1997,s_clarage}. The expression for the diffuse intensity predicted by this model has previously been derived by making use of the Patterson \cite{s_clarage}; here, we provide a derivation based on the scattered intensity in reciprocal space. This model assumes the following:
\begin{enumerate}
\item A global isotropic displacement parameter: $V_{c_i c_i} = \sigma^2$ for all $c$ and $i$.
\item Interatomic covariances are isotropic and depend on interatomic distance: $V_{c_i d_j} = \sigma^2\Gamma(\mathbf{r}_{c_i d_j})$, where the kernel $\Gamma$ describes the correlation length.
\end{enumerate}
With these assumptions, eq.~\ref{eq:mvnscatter} (which still permits correlations between unit cells) can be rewritten:
\begin{equation}
I(\mathbf{q}) = \sum\limits_{c, d} \sum\limits_{i,j} f_i f_j e^{-i \mathbf{q} \cdot \mathbf{r}_{c_i d_j}} 
e^{- q^2 \sigma^2} e^{q^2 \sigma^2 \Gamma(\mathbf{r}_{c_i d_j})}
\label{eq:llmtotal}
\end{equation}
If we additionally assume that $q^2 \sigma^2 \Gamma(\mathbf{r}_{c_i d_j})$ is small, then we can perform Taylor expansion on the exponential $\exp \{ q^2 \sigma^2 \Gamma(\mathbf{r}_{c_i d_j}) \} \approx 1 + q^2 \sigma^2 \Gamma(\mathbf{r}_{c_i d_j})$. Although the exclusion of higher-order terms renders this model less valid at high resolution, this assumption was generally reasonable for the values of $q$ and $\sigma$ considered here. In the case of CypA, for instance, inclusion of the second-order term affected the CC between the predicted and experimental maps by less than 0.01 (data not shown). Eq.~\ref{eq:llmtotal} can then be further simplified:
\begin{align}
I(\mathbf{q}) =&\ e^{- q^2 \sigma^2} \sum\limits_{c, d} 
\sum\limits_{i,j} f_i f_j e^{-i \mathbf{q} \cdot \mathbf{r}_{c_i d_j}} 
\left[ 1 + q^2 \sigma^2 \Gamma(\mathbf{r}_{c_i d_j}) \right] 
\nonumber \\
=&\ e^{- q^2 \sigma^2} \sum\limits_{c,d}
\sum\limits_{i,j} f_i f_j e^{-i \mathbf{q} \cdot \mathbf{r}_{c_i d_j}} 
+ q^2 \sigma^2 e^{- q^2 \sigma^2} 
\sum\limits_{c,d}  
\sum\limits_{i,j} f_i f_j e^{-i \mathbf{q} \cdot \mathbf{r}_{c_i d_j}} \Gamma(\mathbf{r}_{c_i d_j}) \nonumber \\
=&\ e^{- q^2 \sigma^2} I_c 
+ q^2 \sigma^2 e^{- q^2 \sigma^2} \sum\limits_{c,d}
\sum\limits_{i,j} f_i f_j e^{-i \mathbf{q} \cdot \mathbf{r}_{c_i d_j}} \Gamma(\mathbf{r}_{c_i d_j})
\label{eq:llmpartition}
\end{align}
The first term corresponds to the Bragg intensity scaled by a global Debye-Waller factor, while the remaining scattering in the second term corresponds to the diffuse intensity predicted by the liquid-like motions model. This term can be simplified by defining a new kernel function:
\begin{equation}
s(\mathbf{q}, \mathbf{r}) = \sum\limits_{c,d} \sum\limits_{i, j} f_i f_j \delta (\mathbf{r} - \mathbf{r}_{c_i d_j})
\end{equation}
where $\delta(\cdot)$ denotes the Dirac delta function (rather than a displacement). Using this kernel, the second term in eq.~\ref{eq:llmpartition} can be rewritten:
\begin{align}
I_\mathrm{diffuse}(\mathbf{q}) =&\ q^2 \sigma^2 e^{- q^2 \sigma^2} 
\int s( \mathbf{q}, \mathbf{r})  \Gamma(\mathbf{r}) e^{-i \mathbf{q} \cdot \mathbf{r}} \, d \mathbf{r}
\nonumber \\
=&\ q^2 \sigma^2 \, e^{- q^2 \sigma^2} \frac{1}{16 \pi^6} 
\left[ \left( \int s( \mathbf{q}, \mathbf{r})  e^{i \mathbf{q} \cdot \mathbf{r}} \, d \mathbf{r} \right) \ast
\left( \int \Gamma(\mathbf{r})   e^{i \mathbf{q} \cdot \mathbf{r}} \, d \mathbf{r} \right) \right]
\end{align}
where the second step takes advantage of the Fourier convolution theorem, with $\ast$ denoting convolution. Then:
\begin{align}
\int s( \mathbf{q}, \mathbf{r})  e^{i \mathbf{q} \cdot \mathbf{r}} \, d \mathbf{r} = &
\int \sum\limits_{c,d} 
\sum_{i,j} f_i f_j \delta( \mathbf{r} - \mathbf{r}_{c_i d_j}) e^{i \mathbf{q} \cdot \mathbf{r}}  \, d \mathbf{r} 
\nonumber \\ 
=& \sum\limits_{c,d} \sum_{i,j} f_i f_j e^{i \mathbf{q} \cdot \mathbf{r}_{c_i d_j}} = I_c
\end{align}
where $I_c$ is the scattered intensity of the coherently diffracting volume, in this case the crystal transform. The diffuse scattering predicted by the liquid-like motions model is thus a convolution between this intensity function and the Fourier transform of the correlation kernel, $\Gamma$:
\begin{equation}
I_\mathrm{diffuse} (\mathbf{q}) = q^2 \sigma^2 \, e^{- q^2 \sigma^2} \frac{1}{16 \pi^6} 
\left[ I_\mathrm{c} \ast \tilde{\Gamma}(\mathbf{q}) \right].
\label{eq:llm}
\end{equation}
where $\tilde{\Gamma} (\mathbf{q}) = \int \Gamma(\mathbf{r}) e^{i \mathbf{q} \cdot \mathbf{r} } \, d \mathbf{r}$. We call this model, where correlated disorder to extends between neighboring asymmetric units and across unit cell boundaries, the model of liquid-like motions ``with neighbors''.

If, however, we initially assume that correlations are confined within the boundaries of the asymmetric unit ($\Gamma({r}_{c_i d_j}) = 0$ if atoms $c_i$ and $d_j$ are not members of the same asymmetric unit), then we can re-formulate the liquid-like motions model as a function of the molecular, rather than crystal, transform. To see this, start again with the second (diffuse) term in eq.~\ref{eq:llmpartition},
\begin{equation}
I_\mathrm{diffuse} (\mathbf{q}) = 
q^2 \sigma^2 e^{- q^2 \sigma^2} \sum\limits_{c,d}
\sum\limits_{i,j} f_i f_j e^{-i \mathbf{q} \cdot \mathbf{r}_{c_i d_j}} \Gamma(\mathbf{r}_{c_i d_j}) \nonumber
\end{equation}
Under our approximation, many terms in this sum are zero. Specifically, any where atoms $c_i$ and $d_j$ are not members of the same asymmetric unit. Dropping these terms, we obtain
\begin{equation}
I_\mathrm{diffuse} (\mathbf{q}) = 
q^2 \sigma^2 e^{- q^2 \sigma^2} N \sum_{\mathrm{asu}}
\sum\limits_{i,j \in \mathrm{asu}} f_i f_j e^{-i \mathbf{q} \cdot \mathbf{r}_{i j}} \Gamma(\mathbf{r}_{i j})
\label{eq:asuLLM}
\end{equation}
where, again, $\sum_{asu}$ indicates a single summation over all unique copies of the asymmetric unit within a single unit cell replica. Note the sum across unit cells produces the scale factor $N$. No asymmetric unit spans more than one unit cell.

Now we follow the same tactic as above. Define:
\begin{equation}
s_{\mathrm{asu}} (\mathbf{q}, \mathbf{r}) = \sum\limits_{i, j \in \mathrm{asu}} f_i f_j \delta (\mathbf{r} - \mathbf{r}_{i j})
\end{equation}
and re-write eq \ref{eq:asuLLM} 
\begin{align}
I_\mathrm{diffuse} (\mathbf{q}) &= 
q^2 \sigma^2 e^{- q^2 \sigma^2} N \sum_{\mathrm{asu}}
\int s_{\mathrm{asu}} (\mathbf{q}, \mathbf{r})  \, \Gamma(\mathbf{r}) \, e^{-i \mathbf{q} \cdot \mathbf{r}} \, d\mathbf{r}
\label{eq:asuconv} \nonumber \\
&= 
q^2 \sigma^2 e^{- q^2 \sigma^2} \frac{N}{16 \pi^6} 
\sum_{\mathrm{asu}}
\left[ \left( \int s_{\mathrm{asu}} (\mathbf{q}, \mathbf{r}) \, e^{i \mathbf{q} \cdot \mathbf{r}} \, d\mathbf{r} \right)
\ast
\left( \int \Gamma(\mathbf{r}) e^{i \mathbf{q} \cdot \mathbf{r}} \, d \mathbf{r} \right) \right] \nonumber \\
&= 
q^2 \sigma^2 e^{- q^2 \sigma^2} \frac{N}{16 \pi^6} 
\left[ \left( \sum_{\mathrm{asu}}
\sum\limits_{i,j \in \mathrm{asu}}f_i f_j e^{-i \mathbf{q} \cdot \mathbf{r}_{i j}} \right)
\ast
\left( \int \Gamma(\mathbf{r}) e^{i \mathbf{q} \cdot \mathbf{r}} \, d \mathbf{r} \right) \right]
\end{align}
where in the last step we have used the fact that convolution, as a linear operation, distributes. Recalling
\begin{equation*}
\sum_{\mathrm{asu}}
\sum\limits_{i,j \in \mathrm{asu}}f_i f_j e^{-i \mathbf{q} \cdot \mathbf{r}_{i j}} = I_m
\end{equation*}
we obtain 
\begin{equation}
I_\mathrm{diffuse} (\mathbf{q}) = q^2 \sigma^2 \, e^{- q^2 \sigma^2} \frac{N}{16 \pi^6} 
\left[ I_\mathrm{m} \ast \tilde{\Gamma}(\mathbf{q}) \right].
\label{eq:llm}
\end{equation}
The lack of coherence between unit cells results in loss of the Dirac comb from the intensity function, and the addition of a factor that scales the expression by the number of unit cells, $N$. We refer to this as the ASU-confined liquid-like motions model.

For both models, we employed a previously-described form of the kernel, in which covariances decay exponentially as a function of interatomic distance \cite{s_clarage, s_wall1997}:
\begin{equation}
\Gamma(\mathbf{q}) = \frac{8\pi\gamma^3}{(1+q^2\gamma^2)^2}
\label{eq:kernel}
\end{equation}

Prior studies on the liquid-like motions model with neighbors analyzed experimental maps that sampled the observed diffuse scattering intensity at integral Miller indices only \cite{s_wall1997}. In this study, we compare the predicted signal to experimental maps that sample the diffuse signal at fractional Miller indices, which enables us to assess long range correlations that extend beyond a unit cell. For both liquid-like motions models, best fit values of the isotropic displacement parameter and correlation length were determined by a grid search to maximize the CC between the experimental and predicted maps. As an example, convergence of these parameters for the internally-disordered model is shown in Fig. S6B.

\subsection*{Rigid body rotational disorder}
The simplest case of rigid body rotational disorder was evaluated, in which there is no preferred axis of rotation and rotation angles are sampled from a normal distribution. Rotations were additionally assumed to be independent and uncorrelated between asymmetric units. Diffuse scattering maps were predicted from an ensemble of rotated molecules using Guinier's equation \cite{s_guinier}:
\begin{equation}
I_\mathrm{diffuse} (\mathbf{q}) \propto \sum\limits_{\textrm{asu}} \left[ \langle \vert F_{\textrm{asu, n}}(\mathbf{q}) \vert ^2 \rangle - \langle \vert F_{\textrm{asu, n}}(\mathbf{q}) \vert \rangle ^2 \right].
\label{eq:guinier}
\end{equation}
where $F_{\textrm{asu, n}}(\mathbf{q})$ represents the asymmetric unit transform for the $n$th ensemble member and $\langle...\rangle$ indicates the time or ensemble average (which these data cannot distinguish between). In order to focus on the diffuse scattering predicted solely by rotational disorder, the asymmetric unit transforms were not scaled by Debye-Waller factors, which account for translational disorder effects. For each map, the best fit standard deviation of the rotational distribution was determined by scanning over values of this parameter to maximize the CC between the experimental and predicted maps (Fig. S6A). Convergence of the ensembles was determined by ensuring that the CCs with the experimental map were within 0.01 for independent `trajectories' generated with the same rotation parameter.

\subsection*{Ensemble models}
Many types of protein disorder involve transitions between discrete states rather than along a continuum of alternate conformations. In real space, this disorder can be modeled as an ensemble of representative ``snapshots'' of distinct protein configurations. The refined Bragg coordinates of the CypA and AP crystal structures suggested the existence of specific ensembles for these systems. Here we chose the simplest representation of each ensemble: a two-state model, with each state represented by a single, probability-weighted conformation. Probability weights were derived from the refined occupancy values in the Bragg model, and each state was assumed to exhibit uncorrelated atomic disorder that could be adequately described by isotropic B-factors. Diffuse scattering maps were predicted from each two-state ensemble using Guinier's equation (eq. \ref{eq:guinier}) \cite{s_guinier} and scaled by the Debye-Waller factor, $e^{-q^2\sigma^2}$, with the global atomic displacement factor $\sigma$ computed from the Bragg Wilson B factor.  

\clearpage
\section{Extended Figures}

\begin{figure*}[!hbp]
\includegraphics{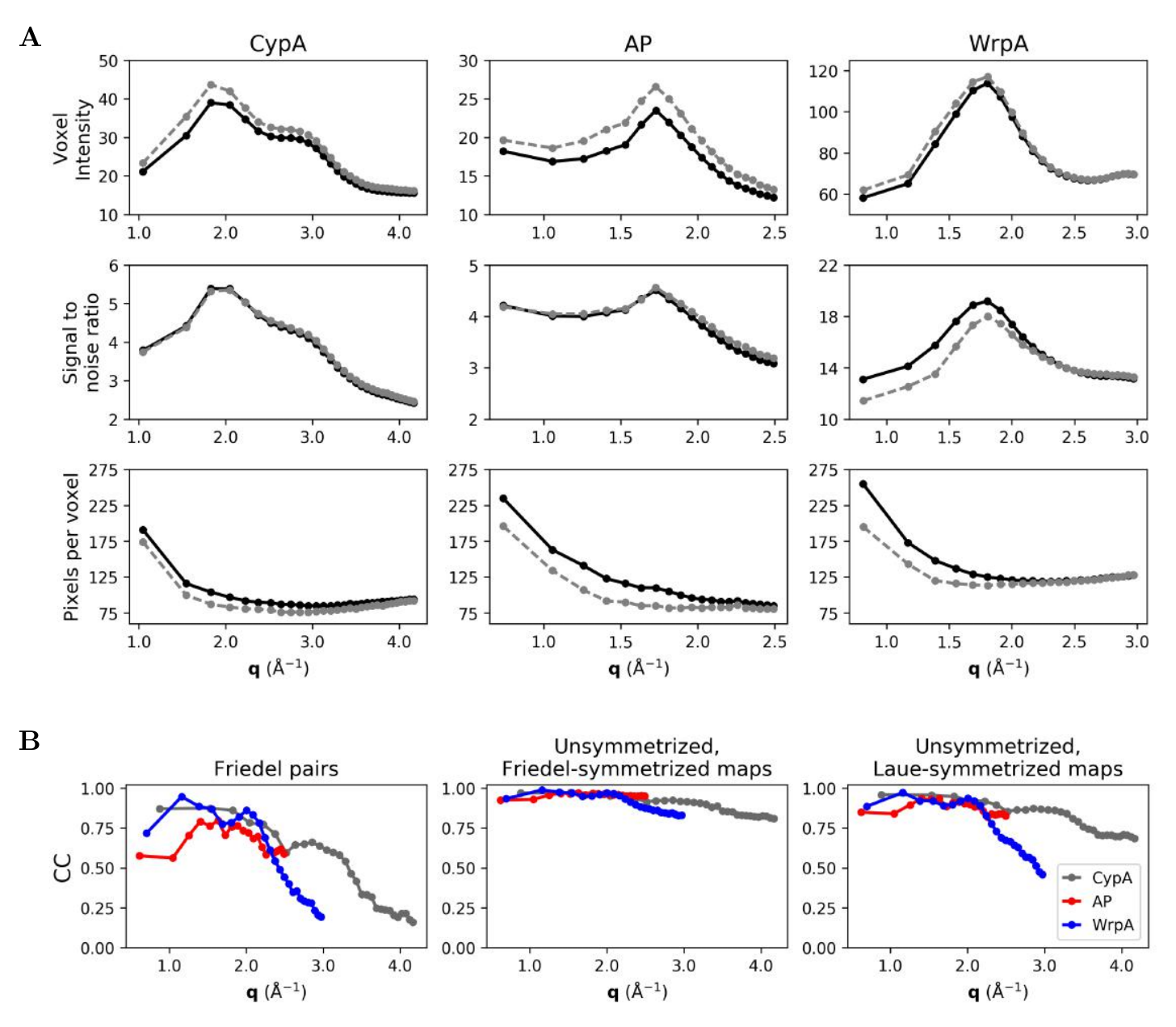}
\centering
\caption{\label{fig:SI_stats}\textbf{Diffuse scattering map statistics by resolution shell.} (A) The median voxel intensity, signal to noise ratio, and number of pixels per voxel are shown across resolution bins for the indicated experimental map. The signal to noise ratio for each voxel was estimated as $\langle I \rangle / \sigma (I)$ for the set of pixel intensities binned into each voxel. The solid and dashed lines correspond to the overall values and the values for voxels centered on integral Miller indices, respectively. (B) Correlation coefficients between Friedel pairs (left), the indicated map before and after averaging Friedel pairs (center), and the indicated map before and after averaging Laue-symmetric voxels (right). }
\end{figure*}

\begin{figure*}[!ht]
\includegraphics[width=.9\linewidth]{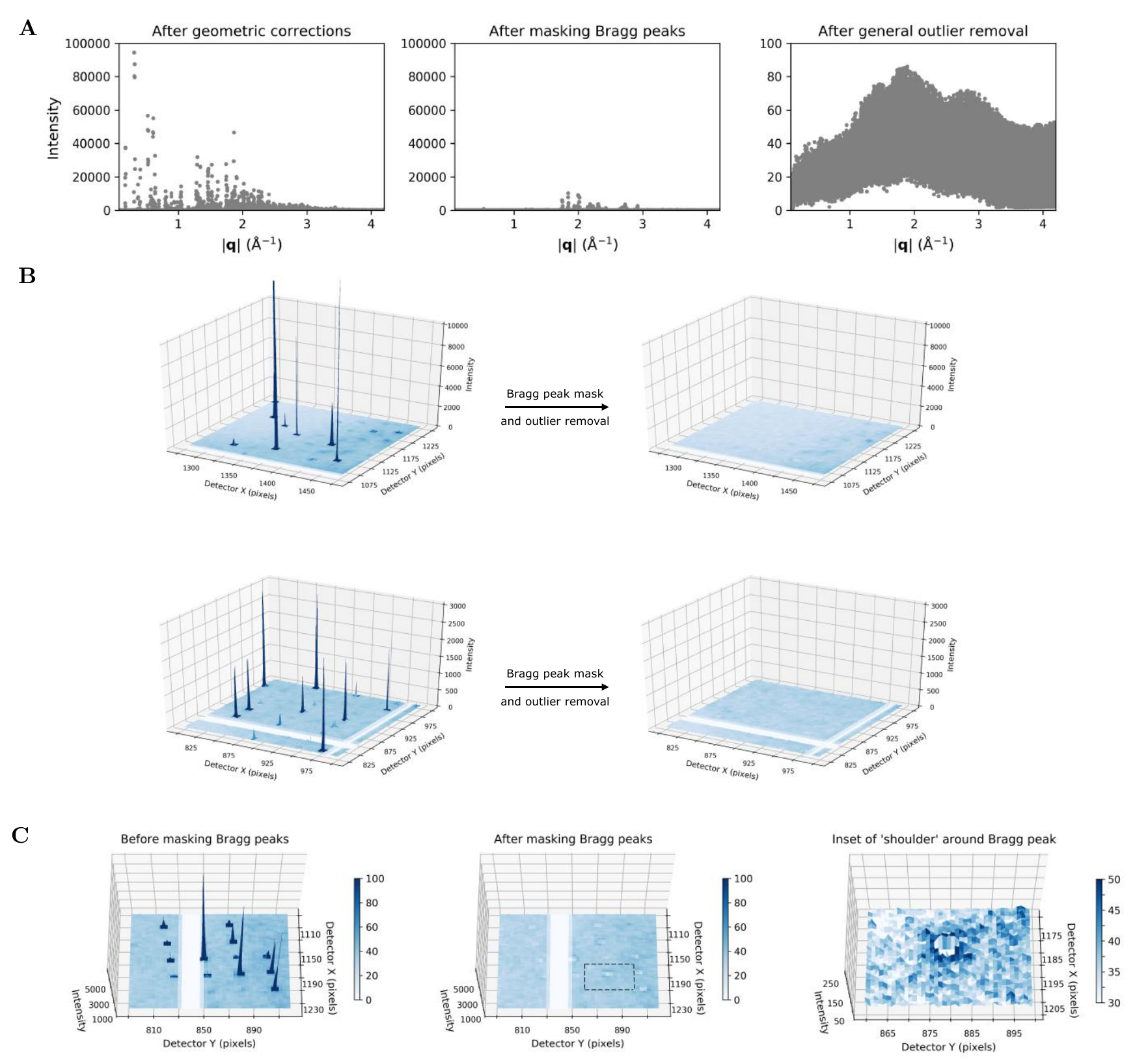}
\centering
\caption{\label{fig:SI_stats}\textbf{Elimination of Bragg peaks.} (A) Intensity distributions for a representative diffraction image from the CypA dataset after applying geometric corrections, masking Bragg peaks, and general outlier removal from left to right. (B) Surface plots of a 200 x 200 pixel region of the image at 5.7 (top) and 2.3 \AA{} (bottom) resolution before and after removal of Bragg peaks and outliers. (C) Representative surface plots as in (B), except tilted. The rightmost panel is shows the diffuse `shoulder’ around the masked Bragg peak boxed in the center panel.}
\end{figure*}

\clearpage
\begin{figure*}
\centering
\includegraphics[width=.9\linewidth]{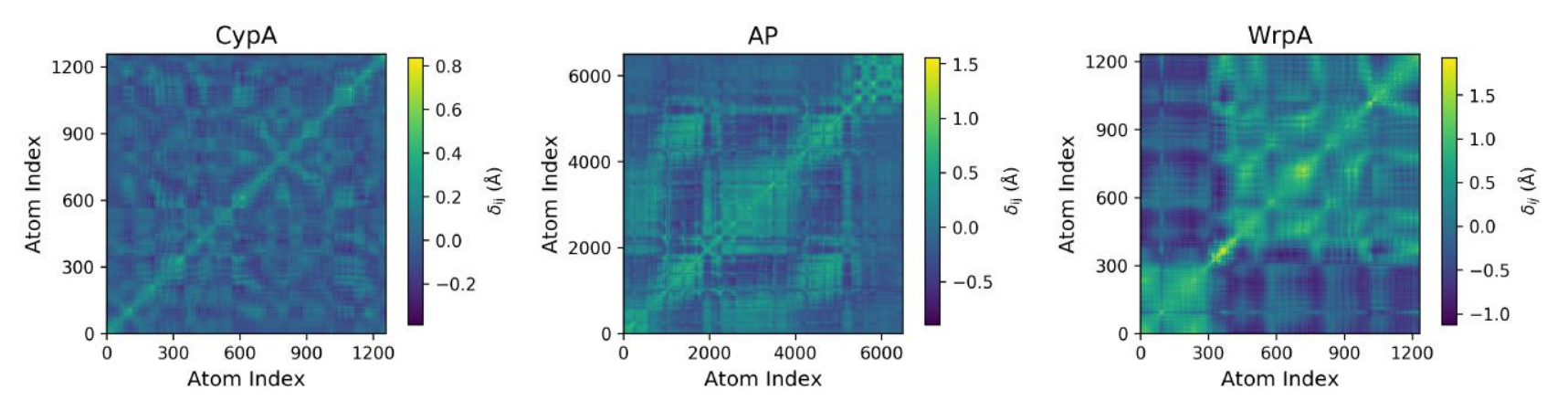}
\caption{\label{fig:SI_stats}\textbf{Covariance matrices generated from elastic network models.} The correlation matrix of interatomic Gaussian displacements was predicted for each system by an elastic network model. The covariance matrix was then computed from this correlation matrix and normalized by the refined B factors to ensure consistency with the Bragg data. }
\end{figure*}

\clearpage
\begin{figure*}
\centering
\includegraphics[width=.9\linewidth]{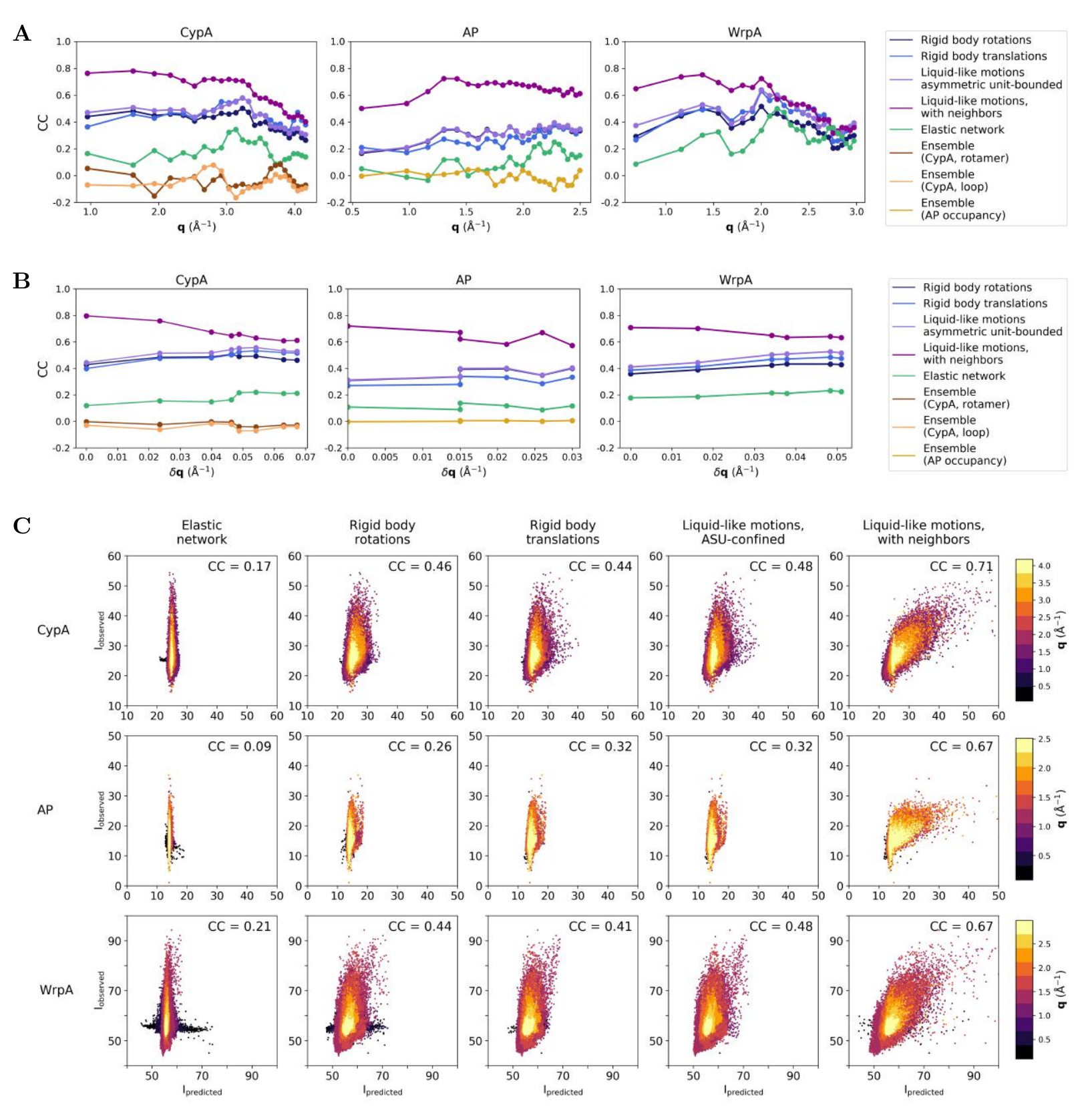}
\caption{\label{fig:SI_stats}\textbf{Correlation coefficients by resolution shell.} The multiplicity-weighted correlation coefficient between the experimental and predicted map for the indicated models is plotted as a function of (A) resolution shell and (B) distance from reciprocal lattice sites. (C) Scatter plots of the model-predicted versus observed intensities, with points colored by resolution; the overall correlation coefficient is noted in the upper right corner of each plot. }
\end{figure*}

\clearpage
\begin{figure*}
\centering
\includegraphics[width=.85\linewidth]{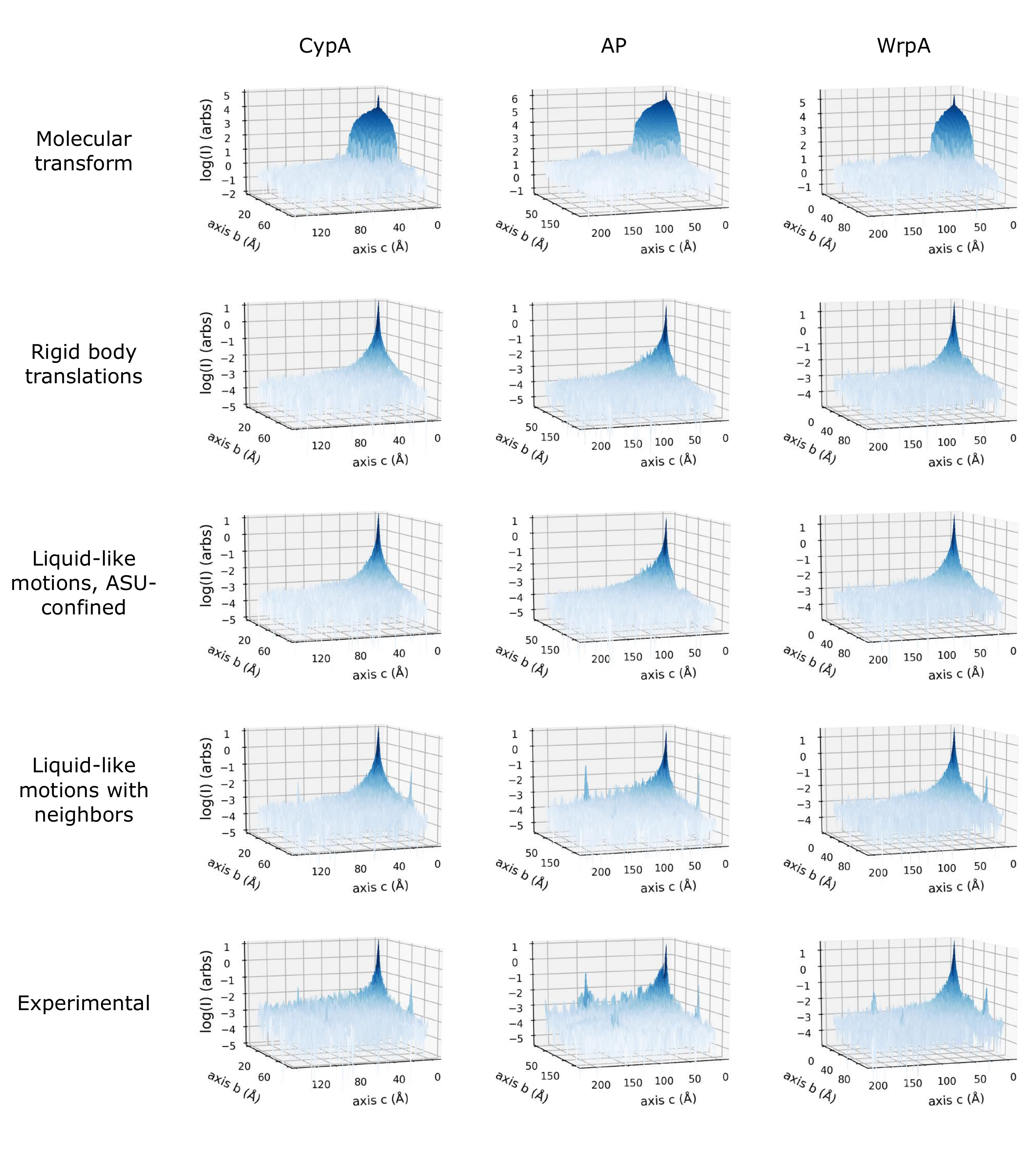}
\caption{\label{fig:SI_acf}\textbf{Autocorrelation functions of the predicted and experimental maps.} The 3D autocorrelation function of the indicated map was computed by Fourier methods; shown are quadrants from the projection along the crystallographic a axis. The low intensity peaks near the boundaries of the map are consistent with the unit cell dimensions along the relevant crystallographic axis. Note that difference vectors in autocorrelation space overestimate the shape of the real space object by a factor of two, but this effect is balanced by only viewing a quadrant of the map (and considering distance to the origin, rather than to the symmetric peak).}
\end{figure*}

\clearpage
\begin{figure*}
\centering
\includegraphics[width=.9\linewidth]{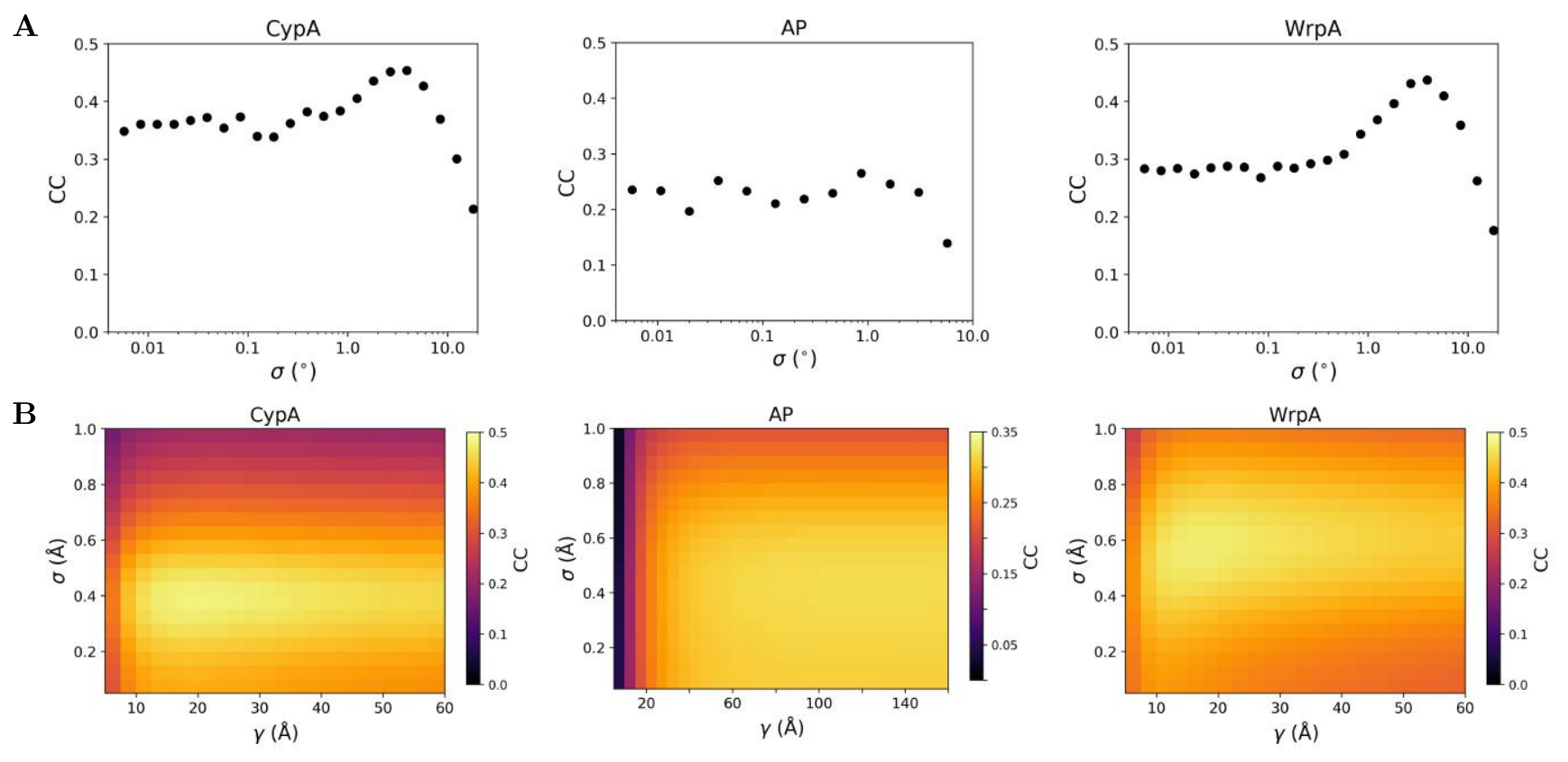}
\caption{\label{fig:SI_stats}\textbf{Convergence of disorder parameters for the rigid body rotational disorder and liquid-like motions models.} (A) Scans over the standard deviation of the rotational distribution, $\sigma$, were performed to fit the rigid body rotational disorder model to each experimental map. (B) Grid scans over the global atomic displacement factor, $\sigma$, and the correlation length, $\gamma$, parameters of the liquid-like motions model. The color indicates the overall correlation coefficient between the experimental and predicted maps. }
\end{figure*}

\end{document}